**Microstructural diversity, nucleation paths and phase behaviour in binary mixtures of charged colloidal spheres**

Nina Lorenz[1], Ishan Gupta[2], Thomas Palberg[1]

[1]Institute of Physics, Johannes Gutenberg University, 55122 Mainz, Germany

[2]Graz University of Technology, Institute of Applied Mechanics, Graz, Austria

We study low-salt, binary aqueous suspensions of charged colloidal spheres of size ratio $\Gamma = 0.57$, number densities below the eutectic number density $n_E$, and number fractions of $p = 1.00\text{-}0.40$. The typical phase obtained by solidification from a homogeneous shear-melt is a substitutional alloy of body centred cubic structure. In strictly gas-tight vials, the polycrystalline solid is stable against melting and further phase transformation for extended times. For comparison, we prepare the same samples also by slow and mechanically undisturbed deionization in commercial slit cells. These cells feature a complex but well reproducible sequence of global and local gradients in salt concentration, number density and composition as induced by successive deionization, phoretic transport and differential settling of the components, respectively. Moreover, they provide an extended bottom surface suitable for heterogeneous nucleation of the β-phase. We give a detailed qualitative characterization of the crystallization processes using imaging and optical microscopy. By contrast to the bulk samples, the initial alloy formation is not volume-filling, and we now observe also α- and β-phases with low solubility of the odd component. In addition to the initial homogeneous nucleation route, the interplay of gradients opens various further crystallization and transformation pathways leading to a great diversity of microstructures. Upon subsequent increase in salt concentration, crystals melt again. Wall-based, pebble-shaped β-phase crystals and facetted α-crystals melt last. Our observations suggest that the substitutional alloys formed in bulk experiments by homogeneous nucleation and subsequent growth are mechanically stable in the absence of solid-fluid interfaces but thermodynamically metastable.

Corresponding author: palberg@uni-mainz.de

**Introduction**

In eutectic phase diagrams, a binary mixture shows a common melting temperature lying well below that of each pure components. The minimal melting temperature, $T_E$, and the corresponding composition, $p_E$ (expressed in terms of the molar ratio $x$, or number fraction $p = n_{small}/n_{total}$), define the eutectic point. In eutectic solidification, the liquid converts simultaneously into two coexisting solid phases. [1]. This is helpful for the processing of eutectic systems. Desired microstructures can be obtained by tuning the temporal sequence of cooling and tempering. The industrially important example of steel forming adjacent crystallites of pure α- and β-phases is well known. During it´s growth at $T < T_E$, particles expelled from one phase feed the growth of crystallites of the other one in their immediate environment, and rod-like or lamella microstructures are quite common. This "in situ"-composite microstructure is highly valued for its wear resistance.

Eutectic phase diagrams also appear in other systems, including colloidal model suspensions. Charged sphere suspensions in particular share a great deal of structural analogies with metallic systems [2]. As in metallic systems, the ratio of thermal energy to interaction energy rules their phase behaviour. However, colloidal systems are processed at constant (ambient) temperature and their interaction strength is varied e.g., by adjusting the number density, $n$. A high colloid density corresponds to a low temperature in a metal system. In low salt ($c \approx 1\mu molL^{-1}$) aqueous suspension, freezing densities are on the order of $n \approx 0.5\text{-}5\mu m^{-3}$ [3, 4, 5]. Corresponding lattice constants are on the order of the wavelength of visible light. On one side, this facilitates structural studies with optical techniques like high-resolution microscopy and Bragg scattering [6]. On the other side, it renders colloidal solids very fragile with shear moduli on the order of a few Pa [7]. This in turn enables access to a homogeneous, gradient-free metastable fluid state simply by shaking the sample [8]. This has been widely used as platform to study the nucleation, growth and coarsening mechanisms and kinetics of colloidal suspensions [9, 10, 11, 12, 13]. Hard spheres [14] and charged spheres of opposite charge form a wealth of compound structures [15]. Concerning like-charged binary mixtures, compound structures were observed e.g. by Hachisu for selected suitable size ratios [16]. Meller and Stavans were the first to report a transition from straight over azeotrope-like to eutectic-like liquidus (L) curves for decreasing size ratio $\Gamma = a_{small}/a_{large}$ (where $a$ denotes the particle radius). Using diffusing wave spectroscopy, the authors further observed a glass phase at intermediate $p$ to separate two crystalline regions at low and at large $p$. Okubo and Fujita [17] investigated exhaustively deionized sys-tems, crystallizing at volume fractions as low as $\Phi = 0.0003$. Also these authors found a systematic dependence on size ratios for L-curves of phase diagrams plotted in the $\Phi_1$-$\Phi_2$ plane. An enhanced crystal stability in a binary mixture of different size but equal charge was reported by Wette et al. [18] and later suggested to result from $AB_4$ compound formation by Liu et al. [19]. Stipp determined the growth velocity of alloy crystals in a binary mixture of size ratio $\Gamma$ = 0.77 and constant $n$ to vary non-monotonously with composition, being lowest at $p \approx 0.8$ [20]. Lorenz et al. investigated both liquidus and solidus curves of binary charged mixtures over a wider range of samples and reported the observation of spindle type phase diagrams and azeotropes [21]. The latter authors also reported the first observation of pure β-phase crystals in a colloidal eutectic forming by differential settling under gravity [22]. These coexisted with the fluid phase and (presumably) small-sphere-enriched alloy crystals. However, in the absence of gradients, samples at $n < n_\mathrm{E}$ solidified completely and formed substitutional alloy phases of composition equal to that of the shear-melt. This is strikingly different to the case of metals, which for $T_\mathrm{L} > T > T_\mathrm{E}$ show the melt to coexist with either α- or β-phase and may raise the question of phase stability. In their seminal work on polydisperse systems, Barrat and Hansen stated the possibility of rapidly forming, mechanically stable crystals, which nevertheless are thermodynamically metastable [23]. Such issues have up to now mainly been addressed by theory and simulation for in hard sphere systems and often in connection with fractionation [24, 25, 26, 27] but not for binary charged sphere mixtures.

Due to their softness, colloidal systems can easily be influenced by external fields to steer their phase behaviour and crystallization path [28]. This either works *via* a manipulation of the particle-particle interaction [29, 30] or *via* an increase in the number density by gravity [31, 32], dielectrically [33, 34] or electrically driven [35], directed transport. Alternatively, the density can also be controlled by hydrodynamic or phoretic transport, e.g., using or micro-pumps [36] or diffusiophoresis in salt concentration gradients [37]. The microstructure of solids and the shape of crystals has been manipulated by mechanical or gravity-induced shear [38, 39, 40, 41] or by local density gradients [42]. Most studies investigated single component systems, much fewer addressed on binary systems [43, 44].

The present study is based on those previous works, and it studies low salt systems at $n < n_\mathrm{E}$ by imaging and optical microscopy. However, it goes beyond in several respects. We conduct a systematic comparison of the solidification behaviour of low salt binary mixtures in dependence on the quench path. Experiments on bulk solidification are repeated with special focus on the previously less well covered region of large $p$. For a second mixture of similar size ratio, they are complemented by experiments on solidification after directed shear. The main part of our new study, however addresses solidification in commercial slit cells. In slit cells, a complex sequence of local and global gradients in salt concentration, number density and composition evolves. In particular, we exploit the *internal* formation of a gradient in interaction strength caused by an ion exchange generated concentration field. The concentration gradient further leads to large scale phoretic transport [45]. Even more important, the two components are influenced differently bay gravity, which leads to differential settling of the large component and creaming of the smaller against the stagnant solvent [46]. Each of these processes has its characteristic time scale, which allows for a complex interplay. Most previous studies addressed systems crystallizing completely via homogeneous nucleation and growth [13]. In slit cells, samples do not crystallize completely, which leaves sufficient room and time for the transport processes to occur. This way we can circumvent the formation of a mechanically stable alloy phase, inhibiting further development towards equilibrium. We observe and characterize a characteristic sequence of observable microstructures and crystal shapes leading from polycrystalline alloys to purified α- and β-phases. In addition, after a few months, a slow increase in electrolyte concentration allows to access the thermodynamic stability of the formed phases and crystal types. Our results demonstrate that the substitutional alloys formed in bulk experiments are indeed mechanically stable but thermodynamically metastable with respect to a three-phase equilibrium. We anticipate that our observations relating specific gradients to characteristic microstructures may aid development of a genuine microstructure-control in colloidal eutectics.

**Experimental**

We studied two binary mixtures of charged sphere species, PnBAPS70-PnBAPS122 and PnBAPS118-PS392. PnBAPS denotes a Poly-n-Butylacrylamid/Polystyrene-copolymer, PS denotes pure Polystyrene. The numbers give the particle diameter in nm as obtained from ultracentrifugation and static light scattering. The copolymer particles were a kind gift of BASF, Ludwigshafen. For the first mixture, the size ratio is $\Gamma = a_{\text{small}}/a_{\text{large}} = 0.557$. From shear modulus measurements the particles carry effective charges of $Z_{\text{G}} = 331\pm3$ and $Z_{\text{G}} = 582\pm18$, respectively. This yields and an interaction-strength-related charge ratio of $\Lambda_{\text{G}} = 0.57$ [21, 22]. Using a mass density of 1.05 g/cm$^3$, their Stokes sedimentation velocities in water are 8.5μms$^{-1}$ and 27.3μms$^{-1}$, respectively. The phase diagrams of the two pure species are given in Fig. S1a and S1b in the SI. The second mixture has a size ratio of $\Gamma = 0.30$ and a charge ratio of $\Lambda_{\text{G}} = 0.31$.

Both systems were prepared in low salt aqueous suspension. Samples were conditioned in different ways and in different cell types. This allows a comparison of crystallization scenarios and resulting microstructures for samples prepared at the same location in the $n$-$p$ phase diagram. After pre-conditioning, samples were either circuit or batch conditioned. In the following we give a short sketch of both and the cell types used with them. For more details, the interested reader is referred to the SI.

Circuit conditioning provides a homogenized shear melt of well-controlled low salt concentration [47]. Samples crystallize from the latter in the absence of any gradients in salt concentration, number density or composition. This approach, in addition facilitates the application of directed shear immediately before the start of the crystallization [48]. The resulting microstructures were observed by microscopy in four-wall-polished flow-through cells with rectangular cross-section of 1mmx10mm (Light Path Optical, Milton Keynes, UK). Samples were mounted to the stage of an inverse microscope (IRB, Leica, Wetzlar). They were observed with 10×, 25×, or 40× magnification in bright field transmission (TM), polarization microscopy (PM) [49], as well as in Bragg microscopy (BM) [50] in transmission (TBM), or reflection under external illumination with parallel white light (RBM) [21]. These cells allow observation in top- and side view, facilitating determination of both microstructure and $n_F$ [53].

In batch-conditioned samples, the systems are in contact with small amounts of ion exchange resin (IEX) in either cylindrical vials or slit cells closed by gas-tight caps. Deionization starts from a salt-free state, equilibrated in contact with ambient air. This state contains a small amount of carbonic acid from dissociation of dissolved $CO_2$ [51] and a negligible amount of cations leached from the cell walls ($\approx 10^{-8}$molL$^{-1}$, [36]). In contact with the IEX, the salt concentration is slowly lowered over time. In return, gradients in salt concentration appear. Crystals appear where- and whenever the electrolyte concentration is lower than the freezing concentration [52]. This has different effects in the vials and in the slit cells. In the vials, it leads to convectional motion facilitated by diffusiophoretic transport and crystal sedimentation. Over the time scale of weeks, also the differential settling of the larger component and creaming of the smaller can be observed. With some extra measures taken, it is, however, well suited for the determination of the phase behaviour. To that end, the vials are occasionally shaken, returning them to a gradient-free melt state. Samples with $n > n_M$, then crystallize in a spatially homogeneous way throughout the vial. Complete deionization is reached, when, after shaking and re-crystallization, the crystallite size has reached a constant value.

Samples prepared in the coexistence region again show more complicated microstructures. However, the vial geometry often inhibits a detailed observation.

Crystallization in vials was followed by visual inspection and photography using ambient or oblique white light illumination and a 12.8Mpx consumer DSRL (D700, Nikon) equipped with a 24-70mm f2.8 zoom lens. If desired, samples in cylindrical vials can be further analysed by conventional static light scattering (SLS) to determine the crystal structure and exact number density as well as their shear modulus [21, 22].

The situation is different for batch-conditioning in slit cells. These feature a 7.5×40mm$^2$ observation region of 250μm height connected at both sides to cylindrical reservoirs, each of approximately 1cm$^3$ volume (Fig. S3a, SI). Samples are imaged by photography and in different microscopy modes. After filling and closing the cells by gas tight screw caps, samples were kept observation region up, vial cap down. The IEX sinks to the bottom of both reservoirs. Slit cells evolve the same gradient types as before. However, now, the observation region remains undisturbed by convection, and it is not mixed again. In particular, the evolving gradient in electrolyte concentration produces an advancing location of the freezing conditions (Fig. S3b, SI). It advances vertically in the reservoirs. In the observation region, the increase is horizontal towards its centre. Slit cells thus translate the temporal sample evolution into a spatial one.

Two further gradient types are observable in slit cells. First, growing crystals with composition close to that of the α- or β-phase tend to exclude the respective odd species, which enriches close to the crystal surface. Second, the continued $CO_2$ leakage at small rates becomes unbalanced by continued ion exchange upon IEX exhaustion. This leads to a secondary gradient in electrolyte concentration. Careful additional cell sealing and use of a sufficient amount of IEX facilitate a shift in the onset of exhaustion towards several months, i.e. beyond the typical time scales of crystallization and microstructural transitions. It then allows melting experiments on the different crystal types.

**Results**

Bulk phase behaviour and micro-structure

Data on the bulk behaviour of the PnBAPS70-PnBAPS122 mixture is displayed in Fig. 1. For densities $n > n_\text{M}(p)$, samples crystallize completely *via* homogeneous nucleation and form a polycrystalline microstructure. The crystallite size increased with decreasing number density well as with decreasing distance to the eutectic composition. Figure 1a shows an example imaged at constant $n = 38 \mu m^{-3}$. Fig.1b shows the phase diagram in the previously less well covered region of $1.0 > p > 0.5$ and $n \leq 100 \mu m^{-3}$. At low densities the sample remains fluid. A coexistence region separates the fluid region from a fully crystalline region. As in previous studies [21, 22], the mixture forms substitutional alloy crystals of body centred cubic (bcc) structure. Samples prepared above the coexistence region showed a very slow increase of crystallite size over the first few months due to coarsening. However, under our gas-tight conditions, they are otherwise stable against further phase transformation or melting for at least four years.

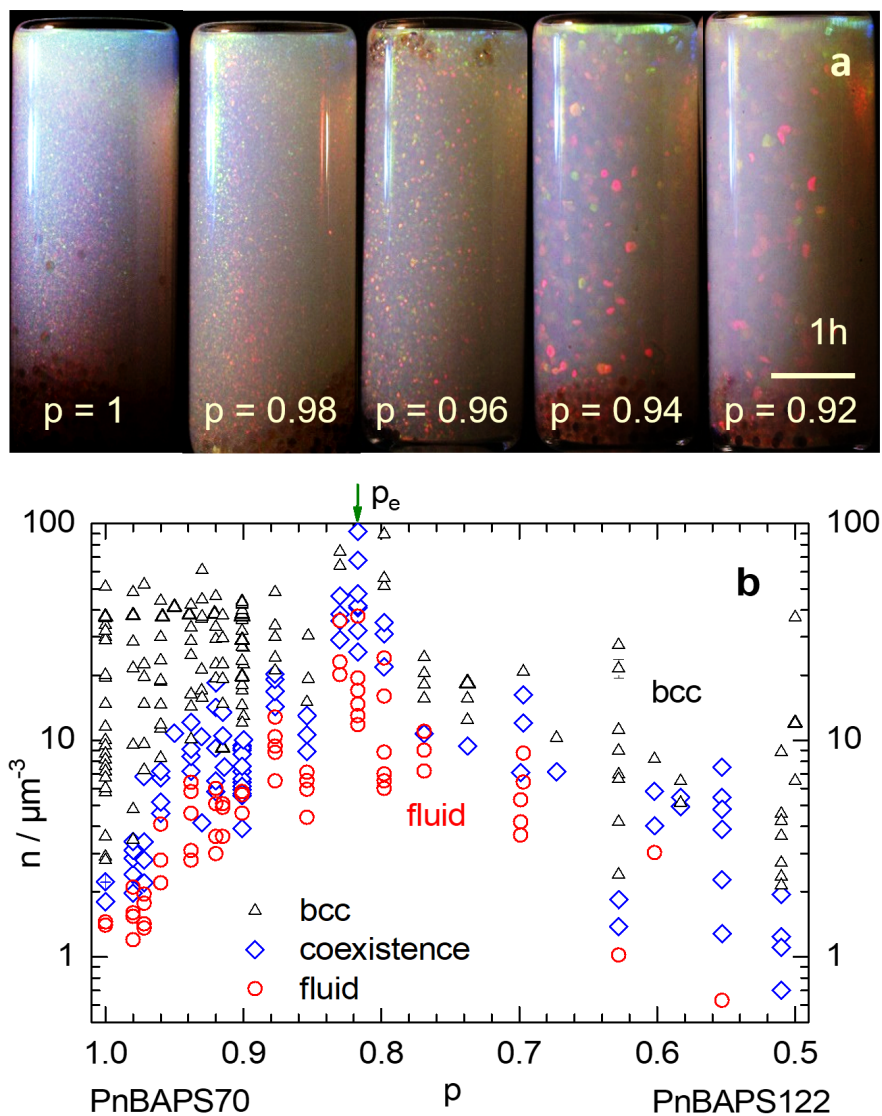

Fig. 1. a) Photographic images of thoroughly deionized bulk samples of the binary mixture PnBaPS70/PnBAPS122 as observed at $t = 1$h after last shaking. Samples were batch conditioned at $n = 38 \mu m^{-3}$ with small-particle number-fractions $p$ as indicated. Scale bar: 500μm. b) Phase diagram as obtained from visual inspection after one month. The green arrow marks the eutectic composition $p_e \approx 0.82$.

In the coexistence region, we observed some microstructural variations. For some representative examples, see Fig. S4 of the SI. Again, we mainly obtained polycrystalline microstructures, but now with a broad spread of crystallite sizes. In several samples, a non-negligible amount of crystals nucleated heterogeneously at the container wall. Quite often, an unambiguous determination of completed crystallization was difficult. Due to this systematic uncertainty, the location of $n_F$ and $n_M$ is not well defined in the phase diagram of Figure 1b. The overall phase diagram shape and the inferred eutectic composition of $p_e \approx 0.82$ (green arrow), however compare well with previous results [21, 22] and expectations from metallurgy [1].

Figure 2 displays bulk data for PnBAPS118/PS392. Here, we focussed on the large $p$-region, $p = (1.00 - 0.96)$ and $n = (0.5 - 10)\mu m^{-3}$, i.e. near the liquidus. For directed shear melting of thoroughly deionized samples, we used circuit conditioning. We observed the samples by microscopy in parallel plate flow through cells. Far above the liquidus, the cell quickly fills with homogeneously nucleated, randomly oriented polycrystals, irrespective of the cycling velocity and the way the shear was stopped (Fig. 2a). Closer to the liquidus, the microstructure becomes shear-history-dependent. After slow cycling and slow stopping, we observe a mosaic of crystals which nucleated heterogeneously at the container walls. In the PM micrograph of Fig. 2b, all crystals show more or less the same blue colour at somewhat differing brightness. This uniformity shows that all crystals are oriented with the most-densely-packed crystallographic plane {110} parallel to the cell wall. The different shading results from random in-plane orientations of <111> within {110} [49]. These quench depth dependent change in nucleation mode are in line with those observed for single component samples [53]. After fast cycling and fast stop of the shear flow, we observe elongated, cloud-shaped, domains of different brightness (Fig. 2c). These relate to the twinning pattern within {110} of the wall-based crystals [54, 48]. For the BM micrograph, the illumination angle was chosen such that only twins with <111> parallel to the formerly applied flow direction yielded observable Bragg reflections [8]. Rather than nucleating in individual events, these crystals emerge *via* a martensitic transition from a thin initial sheet, left-over from the sheared state and of hexagonal order [55]. This microstructure also dominates in circuit conditioned slit cells. (Fig. S5, ESI).

All these microstructures were observed to be stable and did not show further transitions, except for melting upon gas leakage

The phase diagram of the circuit conditioned system in Fig. 2d shows a narrow coexistence region with no ambiguities and only little scatter. The centre of the coexistence region is well approximated by a straight line in this semi-log plot (dashed blue line). For selected samples, we determined the crystal structure by SLS, following [56] (larger symbols in Fig. 2d). Throughout the crystalline region, we observe substitutional alloy crystals. At lower $n$ these are of bcc structure. Note, however, the low-lying transition to face centred cubic (fcc) structure at $n \approx (7\text{--}9)\mu\text{m}^{-3}$ for $p = 1$, and its shift towards larger $n$ with decreasing $p$. Fcc crystals are also recognizable by their striations in the BM micrograph of Fig. 2a [57]. The shear modulus of samples was measured in the bcc region. It varied smoothly with composition with magnitudes in the theoretically expected range of a few Pa.

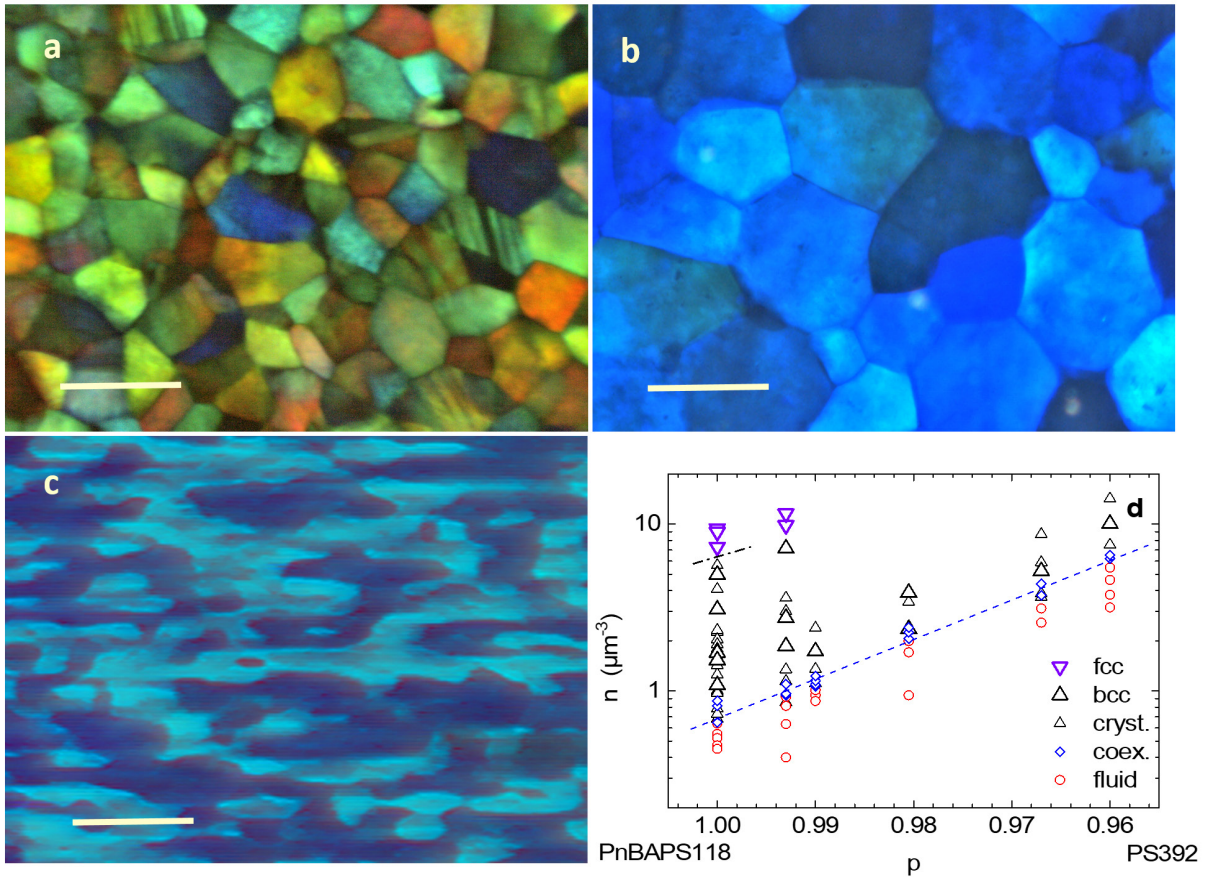

Figure 2: Microstructure and phase behaviour of the thoroughly deionized binary mixture PnBAPS118/PS392 as observed after directed shear melting in a flow through cell. Scale bars 500μm. a) TBM micrograph of randomly oriented polycrystalline material at $p = 0.993$ and $n = 9.8\mu\text{m}^{-3}$. b) PM image of a polycrystalline material with preferred orientation at $p = 0\text{-}967$

and $n = 5.2\mu m^{-3}$. {110} is parallel to the cell wall with random in-plane orientation of <110>. c) RBM image of a twinned bcc sample at $p = 0.96$ and $n = 10.0\mu m^{-3}$ with {110} parallel to the cell wall. Reflected intensity stems only from one of the differing twin orientations. d) Phase diagram in the $n$-$p$ plane. Small symbols denote the phase state determined by microscopy. Large symbols in addition denote crystal structure as determined by SLS.

Crystal formation and transformation in slit cell experiments.

Overview

Slit cells subjected to slow deionization allow to study microstructure evolution in the presence of different gradients evolving over time. In addition, they provide ample opportunity for heterogeneous nucleation at the cell walls. Finally, due to continued slow $CO_2$ leakage and the finite amount of IEX, the electrolyte concentration increases again after several months, which facilitates observations of the relative stability of previously formed microstructures. We focus on the binary mixture PnBAPS70/PnBAPS122 with $n$ chosen a few times larger than $n_F(p)$ (i.e. all these sampled had crystallized completely in the bulk experiments).

Crystallization proceeds in different characteristic stages (labelled I-IV). In each, the dominant crystal types differ in structure, composition and microstructure. This general scenario was followed in practically all our slit cell experiments. However, onset and duration of each stage as well as the location and relative extension of crystal formation type regions depended strongly on the chosen starting conditions ($n_0$, $p_0$). Moreover, the development was hardly ever fully symmetric. This results from the cell-specific differences in gas-tightness and the amount of IEX added to each reservoir. The latter also determines the time span up to which crystals keep forming and transforming. For typical samples, crystal transformations become very slow after 5-6 months and leakage-induced melting commences after some 7-11 months. Fig. 3 shows two representative realizations of the general temporal sequence of crystal formation and the specific alterations in location of formed microstructures. For further examples see Fig. S6, SI.

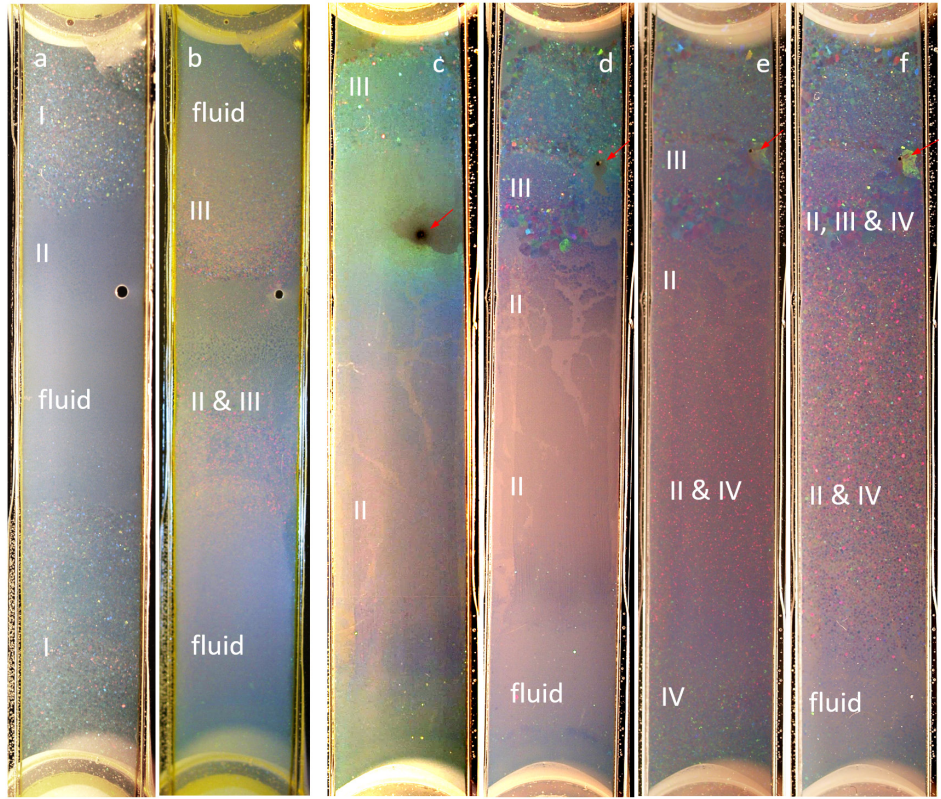

Figure 3. Two representative examples of the spatio-temporal sequence and distribution of crystallization stages for PnBAPS70/PnBAPS122 in slit cells under batch-conditioning. The photographic images were taken under ambient illumination. Image height 51.8mm, image width 7.8mm. The observation area between the reservoir rims is roughly 40×7.5mm$^2$. Roman numerals corresponding to the respective formation stage denote: I: early alloy crystals; II: bottom based, heterogeneously nucleated β-phase crystals; III: Mix of β-phase crystals and small-particle-enriched, heterogeneously nucleated, secondary alloy crystals; IV: late, small-particle-rich, homogeneously nucleated alloy crystals in the top compartment of the cell. The subtle asymmetry of crystal type distribution can be attributed to differing exchange rates in the two reservoirs. a) Sample with $p = 0.4$ and $n = 12.3\mu m^{-3}$ at $t = 4d$. b) the same sample at $t = 19d$. Note the shrinkage of the accidentally present air bubble. c) to f): $p = 0.5$ and $n =12.8\mu m^{-3}$ at different times. c) $t = 7d$; d) $t=15d$; e) $t = 21d$; and d) $t = 35d$. The arrow marks an accidentally present IEX bead. Note its initial drift and the ongoing overall colour changes revealing diffusiophoretic transport processes. Due to this transport, the initially present stage I alloy crystals have already been pushed back into the reservoirs.

Stage I

In the initial crystal formation stage (I) takes a few days. As in the bulk experiments, substitutional alloy crystals form before any significant changes in composition have occurred. A crystallization front advances from the reservoirs towards the cell centre, where the parameters $n$, $c$, and $p_0$ first allow for crystal formation. Its advance is controlled mainly by the eventually flattening gradient in salt concentration. In reaction to the gradient, particles of both kinds are transported by diffusiophoresis away from the centre and towards the IEX placed at the reservoir bottoms. Thus, an additional gradient in $n$ emerges, which leads to a rainbow-like, overall colour change in the fluid part of the images. The observed drift speeds range on the order of cm/week. All samples start crystallizing in the reservoirs, which become deionized first and have the largest $n$. After 1-2 few days, the crystallization front reaches the horizontal observation region. Fig. 3a shows a sample after 4 days. Alloy crystals are readily recognized from their colourful Bragg reflections, they fill roughly half the observation volume. Due to the ongoing transport processes, the fronts are often considerably curved (Fig. S7, ESI).

Stage II

The first stage is ended and the second stage induced by the differential settling of the larger component. This establishes a vertical gradient in composition within the remaining melt. At the cell bottom $p < p_0$. The local decrease in $p$ stops further formation of early alloy crystals, and the formation of nearly pure large component particles (β-phase) commences in the formerly fluid region. These appear first as dark sheets, later transforming to isolated dark spots in the central regions of Fig. 3b to 3e. Typically, we observe a lateral decrease of coverage by β-phase towards the cell centre. Pebbles are thicker than sheets, and sheets gradually thin out and finally disappear.

Stage III

From now on the situation gets considerably more complicated. Stage III starts a few day after Stage II. It is characterized by a vanishing horizontal gradient in salt concentration and a growing vertical gradient in concentration. Several processes now occur at different locations simultaneously. Next to the growing dark spots, a second type of alloy crystals nucleates and grows at the bottom of the observation region, either on top of β-phase crystals or in between. These crystals, again, show vivid Bragg scattering. The region of onset advances towards the centre (c.f. the change in relative position of the crystallization front with respect to the IEX bead marked by the arrow in Fig. 3c and 3d). Both crystal types expel particles of the odd component and enrich them in their immediate vicinity. Thus, in this stage, also local variations in $p$ control alloy crystal formation as well as further β-phase growth. Meanwhile however, the initial alloy crystals disappear. Fig. 3b shows the situation after 19d, i.e. about a week after stage III had started. Secondary alloy crystals and β-phase fill the central region, while most alloy crystals close to the reservoirs have melted. This gave additional room for further and secondary alloy formation. In the top region of Fig. 3b, the initial alloys have already been replaced by alloy and β-phase crystals. Throughout stage III, the dissolution and reformation of alloy crystals continues, giving rise to interesting microstructural details. In most cases it was only stopped after several months by overall melting due to the gradual increase in salt concentration. However, in optimally sealed cells, it may end much earlier (Fig. S6 a-c, SI)

Stage IV

In many samples, the onset of a further stage (IV) could be observed after some 2 to 5 weeks. Here, the smaller component creams upward and fills the upper part of the observation volume. In samples forming an extended fluid region close to and in the top part of the reservoirs during stage III, creaming may even lead to an inversion of the initial density gradient formed by diffusiophoresis (Fig. S6 d and e, SI) In particular for samples with $p_0 > 0.85$, we observe the homogeneous nucleation of alloy crystals in the upper region of the observation slit (Fig. 3e and 3f). Comparison to the crystals formed in bulk samples prepared at coexistence and the corresponding "head"-crystals of Lorenz [22], suggests to interpret these stage IV crystals as alloy crystals forming of significantly enlarged $p$ and $n$. To summarize the main findings of this overview, in slit cells, crystallization proceeds in a fairly well-ordered sequence of different spatio-temporal crystal formation stages.

Specific microstructures

Stage I alloy crystals show up to four different microstructure variants, again forming in a characteristic sequence. Figure 4 shows a representative PM micrograph I taken after 1d. To the left, i.e. close to the reservoir, one finds small crystals of near equal size resulting from an narrow nucleation zone propagating towards the cell centre. Here, the initial burst of homogeneous nucleation followed by growth and intersection (a). As the nucleation front advanced, a second variant of homogeneously nucleated crystals appears (b). These are larger and show pronouncedly curved boundaries (arrows). According to Johnson and Mehl [58], this to an enduring but slow steady state nucleation at moderate metastability combined with growth at constant velocity. The third microstructure variant is not always observed. In Fig. 4, it is identified from comparison to the results shown in Fig. 2b as heterogeneously nucleated, upward growing columnar crystals at the cell wall (c). Finally, columnar crystals nucleate at type-b or type-c crystals (secondary nucleation) and grow in a horizontal direction. (d). Their multi-coloured appearance indicates a random orientation of the crystallite axes. Presumably, differential settling has led to a coverage of the cell bottom by large particles, inhibiting further heterogeneous nucleation of the small-component-rich alloy crystals due to crystal lattice mismatch.

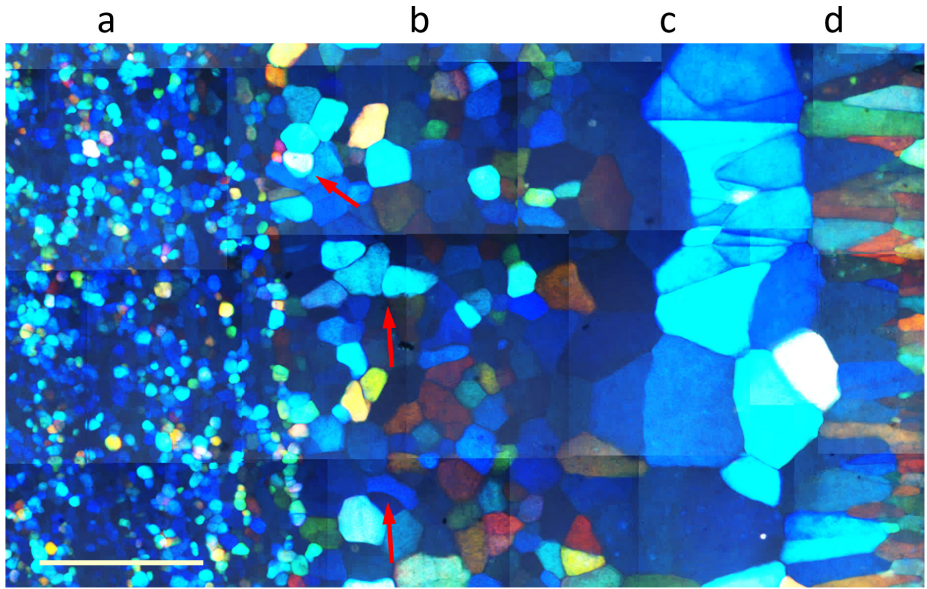

Figure 4: Microstructures of stage I alloy crystals. Collage of polarization microscopic images taken on a sample at $p = 0.96$ and $n = 38\mu m^{-3}$ after 1d. Scale bar 500μm. Different microstructures (labelled a-d) have formed from left to right (for details see text). (a) is about two mm off the reservoir rim to the left. Arrows in zone (b) highlight curved crystal boundaries. Note the absence of any pronounced texturing, indicating that these crystals fill the complete cell height.

β-phase material forms from PnBAPS122 particles, which have differentially settled under gravity and accumulated at the lower cell wall (stage II) or are expelled from growing, secondary alloy crystals (stage III). In the Bragg image of a sample at $p = 0.90$ and $n = 20\mu m^{-3}$ taken after 20d (Fig. 5a), we observe the stage II crystals to the right towards the cell centre and the stage III crystals to the left.

β-phase material can be discriminated from alloy crystals by its distinctive optical properties. In Bragg imaging, it shows reflections only under grazing incidence illumination and observation [22]. Under standard illumination, it appears as black, i.e. non scattering, regions embedded in weakly scattering melt or next to alloy phase crystals which do show Bragg scattering if oriented favourably. With increasing crystal height, β-phase crystals appear ever darker. In Fig. 5b, the stage III region of Fig. 5a is observed again in PM. Now, the β-phase shows only a very faint, blueish hue (Fig. 5b, right). The recorded intensity is orders of magnitude weaker than for alloy crystals of comparable thickness (Fig. 5b, centre and left). Only in regions with no alloy crystals around, the pebble shaped β-phase crystals can be imaged better using long exposure times (Figure 5c).

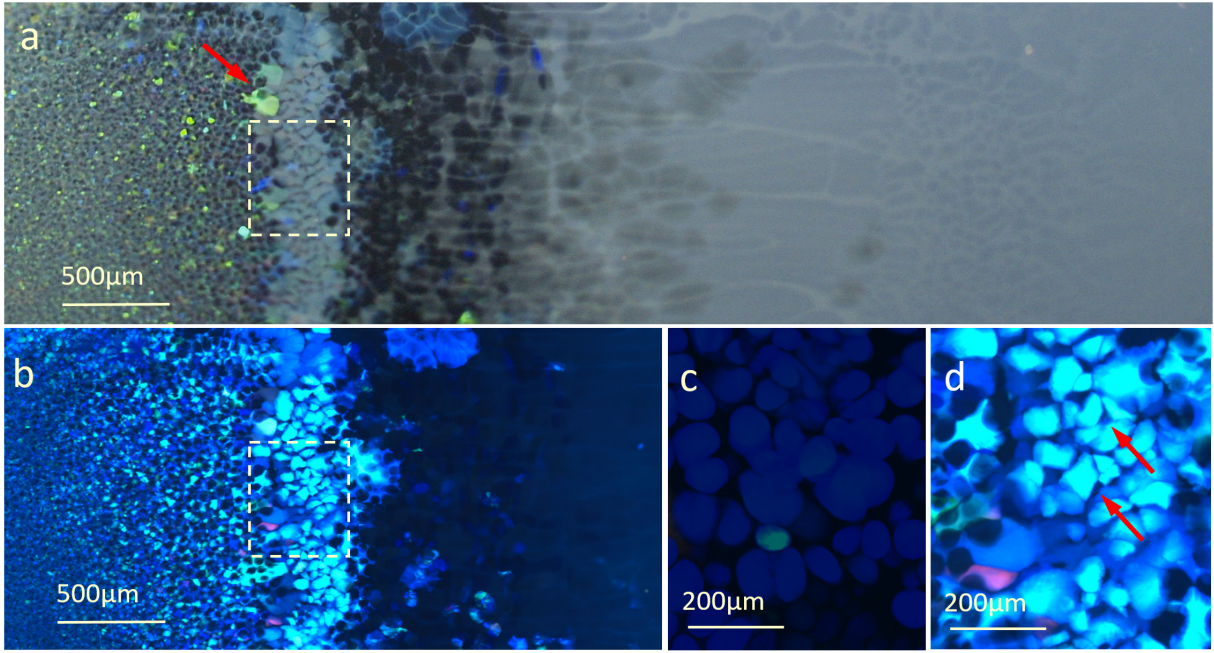

Figure 5: Microstructures of stage II β-phase crystals. The cell centre is to the right, the nearest reservoir to the left. Scale bars as indicated. a) Transmission Bragg image of a sample at $p = 0.90$ and $n = 20\mu m^{-3}$ taken after 20d. A few left-over columnar stage I crystals mark the largest rightward extension of the initial stage I region of alloy crystal formation (red arrow). b) The same sample as observed polarization microscopy. c) PM micrograph of β-phase pebbles close to the centre region of the cell. d) close-up of the marked region in b) showing partially overgrown β-phase pebbles and β-phase confined to the interstitials of stage III alloy crystals (arrows).

Starting a small distance away from the rightmost alloy crystals, the β-phase initially appears as thin, wall nucleated spots of crystalline material which merge to sheets intersected by thin channels (left region in Fig. 5a). The covered region expands and advances with time towards the still fluid cell. Next, the sheets transform into small, half-dome-shaped crystals of roughly circular base as seen in the dark centre region of Fig. 5a, taken after 10d, i.e. after commence of stage III to the left. During this stage III, β-phase grows in concert with alloy crystals. A light gray region fills the previously left gap between alloy and β-phase material. In Fig. 5b this region appears as bright region of blue colour. The highlighted rectangular area in Figs. 6a and b is shown enlarged in Fig. 5d. We observe freshly formed, bottom based alloy crystals with rather straight edges to confine the β-phase between their borders. Still further to the left, isolated β-phase pebble crystals and interspersed secondary alloy crystals grow and replace the vanished stage I alloy crystals. Also here, β-phase material and alloy crystals grow side by side each feeding the other with expelled odd particles, such that locally, $p > p_F(n)$ for both phases. In the metal case, this is known as sympathetic nucleation and growth and responsible for the lamellae-type microstructure observed after solidification at large metastability. In the present experiments, we are at weak metastability ($n < n_E$) and stage II crystals never get space filling, although their final height, as inferred from their relative darkness, increases with $n$.

The second round of alloy crystal formation proceeds in parallel in the region where the stage I alloys had disappeared and towards the central region. In the former small crystals grow next to β-phase pebbles, either on top or in between. Different crystal shapes and microstructure emerge in the central region. Fig. 6 shows a time series of these stage III alloy crystals. (Note the flipped display direction, the IEX is now towards the bottom of the images). In the formerly pebble-rich region (black stripe in Fig. 5a) relatively small alloy crystals grow side by side with β-phase material. Further towards the centre, we see many alloy crystals nucleate and grow on top of the thin β-phase sheets and patches. Interestingly, practically all of them appear to be oriented with a dense packed plane parallel to their substrate and only rarely a differently oriented crystal emerges (yellow arrow in Fig. 6). Neighbouring crystals form facets upon intersection. Crystals growing more isolated, start with roundish shapes but soon develop a pronounced growth faceting. We measured their maximum height by repeated focussing experiments. The sharpness of top surface details allowed to infer an increase of final height from 20-30μm at $n = 15$μm$^{-3}$, over 40-50μm at $n = 20$μm$^{-3}$ to ≈ 70-90μm at $n = 28$μm$^{-3}$. Thus, these crystals stay thin but grow to considerable lateral extension. Also the relative amount of these different microstructures depends on the overall density. We observe a clear trend to increase with overall density and a pronounced scatter from sample to sample.

Here the majority of low heights was observed in samples, which before developed a steep gradient in $c$. Also the number of sympathetically growing columnar alloys and β-phase crystals as well as of isolated β-phase pebbles increased with increasing $n$ (Fig. S8 and S9 of the SI). At low $n$, the thin, growth facetted alloy crystals prevail. Further, at the end of their initial growth period, columnar crystals nucleate at their rim and grow outward. As before in stage I, these appear to have a more random orientation. Fig. 6 shows that this process occurs in a large region simultaneously and, thus, is systematic. However, it presently remains to be rationalized.

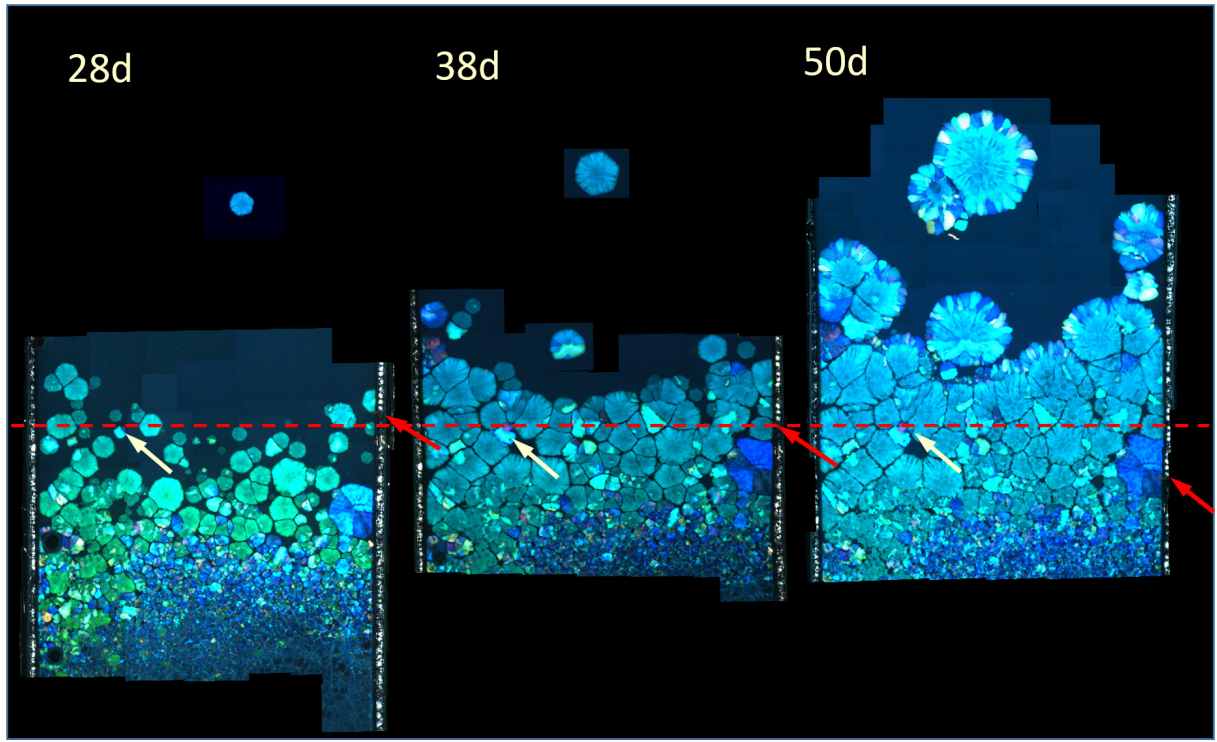

Figure 6: Collages of PM micrographs showing a full cross section of the cell (width 7.5mm) for a sample at sample at $p = 0.94$, and $n = 15$ at different times as indicated. The IEX is towards the bottom, the cell centre towards the top of the collages. The collages are shifted with respect to each other, in order to compensate for the overall diffusiophoretic drift and to display a particular crystal (yellow arrows) at same figure-height (dashed horizontal line). For comparison, the red arrows mark a distinct spot on the cell seam. The change in colour between the first two images results from a slight difference in the orientation of the crossed polarizers. An enlarged version of the $t = 50$d collage can be found in Fig. S8 of the ESI.

This single crystal type shows a rich inner structure or texture. We display some examples in Fig. 7 A typical feature are fine dark lines running outward from the crystal centre (Fig. 7a). These may get more pronounced, with increasing crystal size. Where they reach the crystal edge, one often observes a little kink in the otherwise straight crystal boundary (Fig. 7b). In loose analogy to surface water on glaciers, we attribute them to rivets of expelled large particles collected in scars and flowing outward. The crystal in Fig. 7b also shows roundish, slightly darker features, which result from overgrown flat β-phase patches (see also Fig. S10, SI). Upon encountering more elevated β-phase pebbles, the alloy crystal grows around (Fig. 7c).

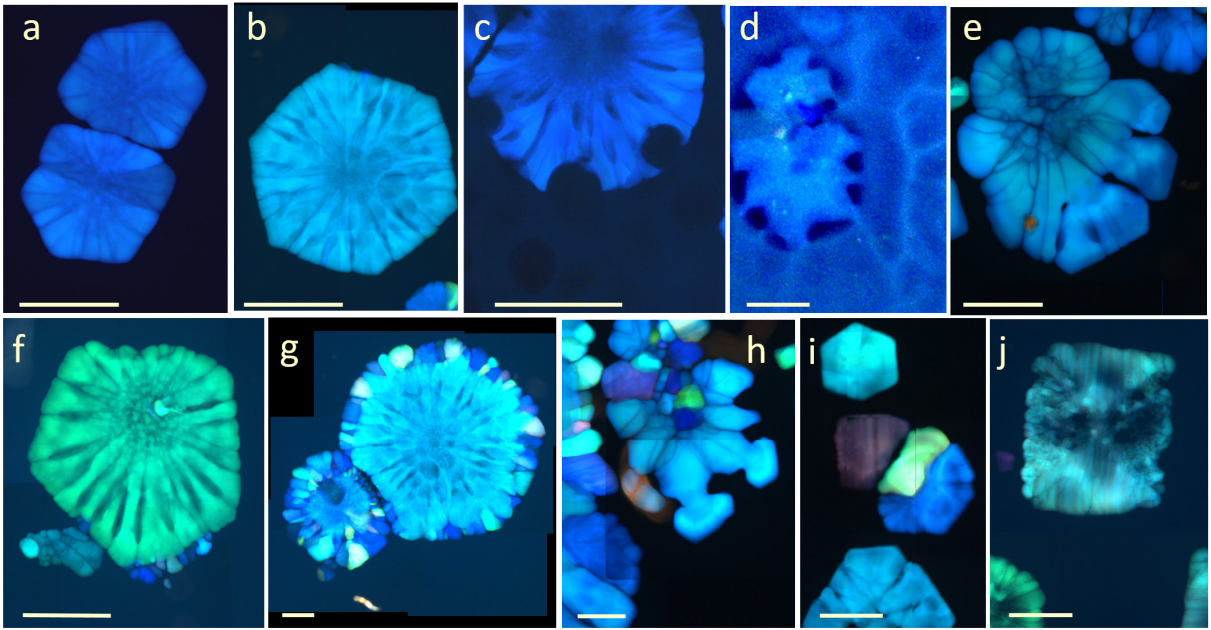

Figure 7: PM close-ups of crystal shapes observable during stage III towards the centre region of different samples. a) to e): samples at $p = 0.94$, and $n = 20 \mu m^{-3}$. In d) we focussed on the cell bottom, to image the β-phase distribution. f) $p = 0.92$, $n = 15\ \mu m^{-3}$ and $t = 48d$. g) $p = 0.94$, $n = 15\ \mu m^{-3}$. Note that this image taken at $t = 50d$ shows the same crystal observed in b at $t = 38d$, but is rotated clockwise by 90°. h) $p = 0.94$, $n = 28\ \mu m^{-3}$ and $t = 42d$. i) $p = 0.94$, $n = 28\ \mu m^{-3}$ and $t = 42d$. j) $p = 0.96$, $n = 12.4 \mu m^{-3}$ and $t = 112d$.

In Fig. 7d, we focussed on the bottom plate to highlight the distribution of large particles. The image allows for two different interpretations. We favour an accumulation of PnBAPS122 at the alloy crystal edge after descending the central plateau. This would be a very slow version of sympathetic growth. Alternatively, Fig. 7d is compatible with the alloy crystal scraping up material from the surrounding β-phase sheet. In particular the correlation of the dark black spots with the surrounding sheet distribution would support this view. At later growth stages, the initially fine black lines become significantly more pronounced. In addition, the initial edge-kinks get deeper and in many cases fjord-like morphologies form, in which the alloy crystals finally form radially oriented, longish inclusions of β-phase (Fig. 7e and f). Faceting at these stages yields more compact crystal boundaries (Fig. 7e). Symmetries are kept across fjords. It is tempting to suspect an increased surface tension of the single crystals towards the adjacent large particle enriched fluid. A close-up of the nucleation and the growth stage of the outer columnar crystals is given in Figs. 7f and g. At larger density, secondary nucleation is also observed on top of this crystal type (Fig. 7h). Also, isolated small crystals are observed more often. These are facetted right from their nucleation stage. Interestingly, their symmetries are sometimes compatible with an fcc structure, sometimes with a bcc structure. In Fig. 7i, the two variants grow in immediate vicinity to each other. Finally, the symmetry of faceting depends on time. In early grown alloy crystals, hexagonal facets dominate. After about two months (i.e. way into stage III and often after the onset of significant creaming (see below)) the relative fraction of rectangular shaped facets increases. A particularly nice specimen nucleated at about 90d and observed after 112d is shown in Fig. 7j.

A final well-defined stage of crystal formation accompanies the ongoing processes of stage II and III after some weeks. A representative example is shown in Fig. 8a to 8c, where throughout the complete observation region, a spatially uniform burst of homogeneous nucleation occurs, followed by very slow growth (see also Fig. 5e and f). The growth velocity here is on the order of a few microns per day. Randomly oriented, vividly Bragg scattering crystals form and stay in the upper compartment. Their size is narrowly distributed and rarely exceeds 100μm. Their intersections are fairly straight. Size is limited both by mutual intersection and simple stop of growth. Thus, by contrast to the polycrystalline stage I material in Fig. 4 and the crystal-creaming-induced compactification in bulk sample (Fig. S4f, SI) we observe a loose, crystal-aggregate-like microstructure, with fluid regions in between individual crystal groupings.

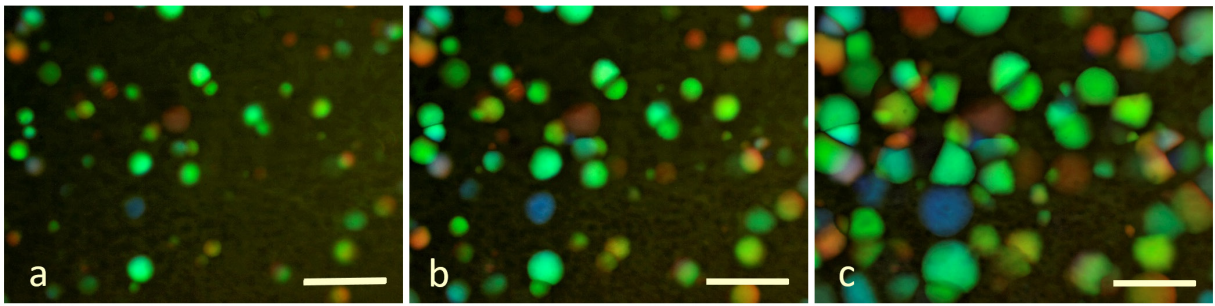

Figure 8: Growth of randomly oriented stage IV alloy crystals nucleated at $t = 28$d in a mixture at $p = 0.5$ and $n = 41.2$μm$^{-3}$. The scale bar is 200μm. The Bragg micrographs were taken under oblique illumination at a) $t = 29$d, b) $t = 31$d, and c) $t = 33$d. Note that practically no new crystals nucleate.

In slit cells, stage IV crystals form most often observed in samples of medium $p$ as well as in samples developing a strong initial diffusiophoretic particle transport towards the IEX. In both sample types, the relative volume of fluid phase is rather large such that large scale creaming can occur. On the other side, mixtures with very low initial $p$ may show creaming, but the concentration of PnBAPS70 stays too low to facilitate nucleation (Fig. S6d and e, SI). Following Lorenz [22], we therefore suggest this stage IV crystals to be PnBaPS70-enriched alloys of $p > p_0$.

Thermodynamic stability.

Throughout stage III elder alloy crystals dissolve and are replaced freshly nucleated ones. The process however slows with time and hardly any fresh crystals form after three to four months. From visual inspection, transformation appeared to continue very slowly only in the reservoirs. In the observation region, stage IV crystals dissolve after three to four months. β-phase sheets, and pebbles, as well as late-formed stage III alloy crystals are fairly stable and regularly survive up to the onset of salt-induced melting.

Slit cells are not completely gas-tight. $CO_2$ leaks very slowly but steadily through their seams and distributes diffusively. Combined with the limited capacity of the IEX this leads to a slow, nearly gradient free increase in salt concentration after some 7 to 10 months, i.e. at times, when crystallization and transformation processes have practically ceased in the observation region. In addition to increased screening, $CO_2$ and its reaction products also lower the surface charge of particles [51]. Due to the lowered interaction strength, crystals start to dissolve. Like the stage I and IV alloy crystals before, β-phase crystals and compact stage III alloy crystals melt from their outside inward upon increasing the salt-concentration. The flat stage III crystals in the cell centre melt start melting in their centre. This is illustrated in Fig. 9. This inverted melting direction is closely correlated with the age of the crystalline material. As a rule of thumb, the later any part of an alloy crystal has grown, the later it starts melting.

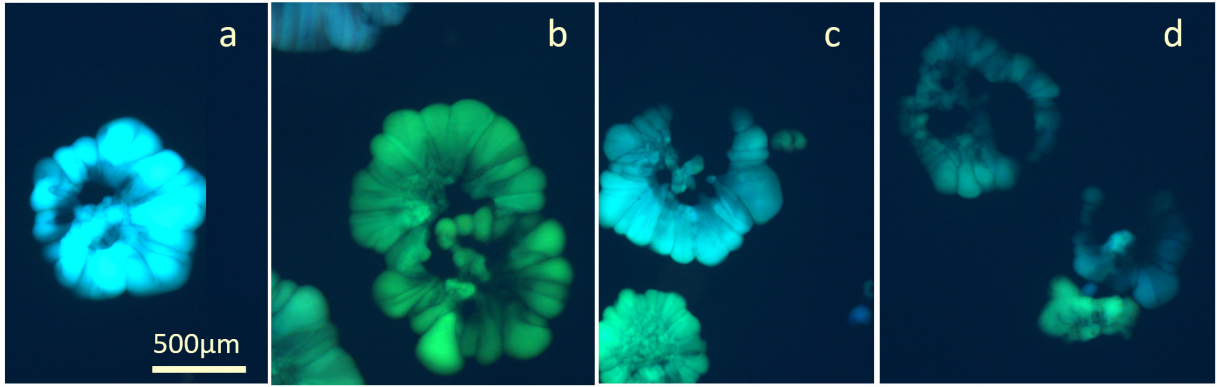

Figure 9: melting from inside out in the flat, facetted alloy crystals grown in the centre. PM micrographs of crystals melting in a sample at $p = 0.94$ and $n = 20 \mu m^{-3}$. Note in c) that the earlier nucleated top crystal has already melted in its central region, while the lower, later nucleated crystal is still intact.

Further we observe a regular sequence of salt-induced melting. It is drawn schematically in Fig. 10a. The last crystals types remaining are late-grown, partially hollowed single crystals (c.f. Fig. 9) occasionally accompanied by overgrown β-phase pebbles and sheets. The relative stability of β-phase crystals is interesting, as these were amongst the first crystal types to appear. The Bragg image in Fig. 11a shows an experimental example taken 8 months after cell filling.

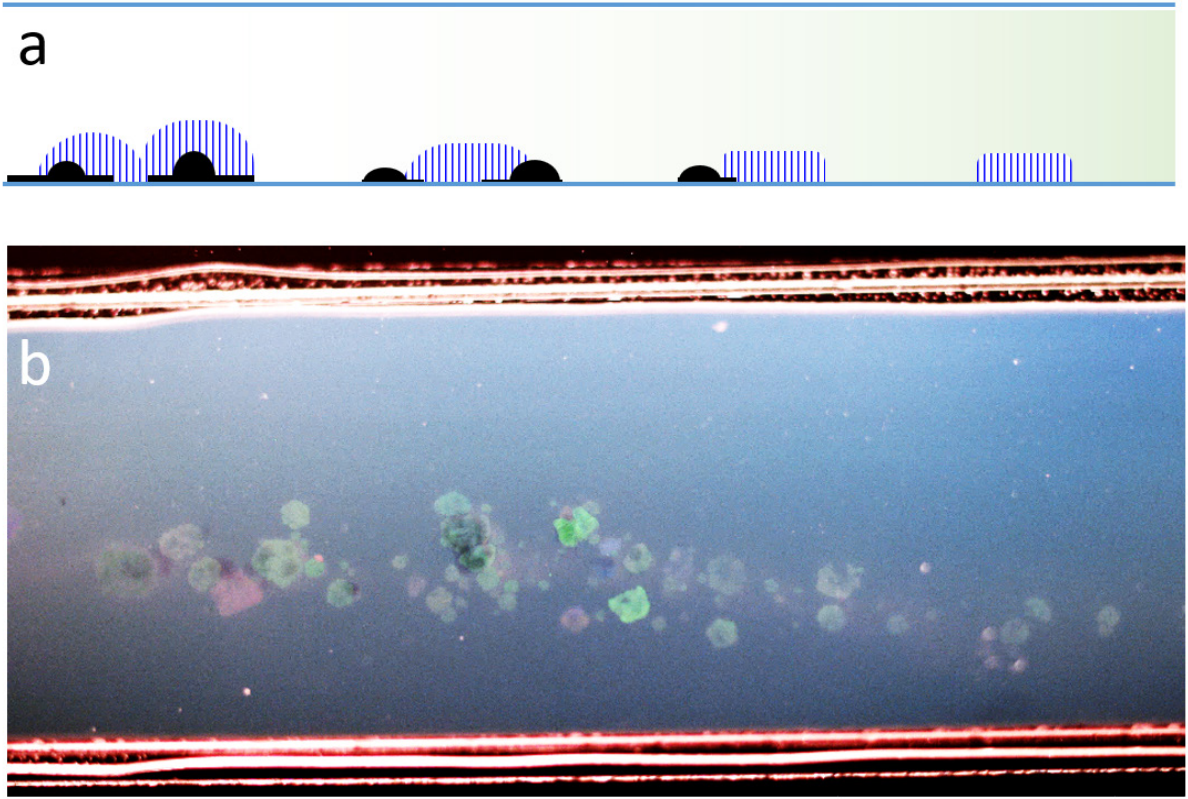

Figure 10: Salt-induced melting. a) Schematic drawing of the sequence of crystal melting and disappearance upon increasing the salt concentration, as indicated by the depth of green shading. Time proceeds from left to right. The underlying salt concentration increases with time. Blue: late grown alloy crystals, black: β-phase crystals. b) Photographic Bragg image of a sample at p = 0.94 and n = 20μm$^{-3}$ taken 8 months after cell filling.

## Discussion

We have presented a comprehensive overview on the phase behaviour, crystallization paths and resulting microstructures for a binary eutectic mixture. We compared crystallization from de-ionized, de-carbonized shear melts (with and without directed shear applied for melting) in gas-tight bulk cells (reference state) to crystallization in slit cells, which started from de-ionized states equilibrated to ambient air. Within the former systems, we regularly obtained polycrystalline substitutional alloys (with some variation due to shear history). In the slit cells, the samples de-carbonize slowly. Thus, a sequence of mutually orthogonal gradients evolves with characteristic times. Global gradients in salt and particle concentration evolved horizontally, while gradients in composition evolved vertically. In addition, local gradients in composition were found to evolve in the vicinity of crystallizing material. We observed characteristic crystallization and microstructural conversion sequences, which were highly correlated to these different gradients and their evolution. Four main stages were discriminated, each characterized by the dominance of specific gradients. After some 4-6 months practically all further microstructure evolution ceased, after 5-7 months the electrolyte concentration increased again due to $CO_2$ leakage. This facilitated the observation of a sequence for salt-induced melting. Slit cells thus allowed to observe a wealth of novel micro-structures not known from bulk crystallization. In the following we will discuss some of these issues under a broader perspective.

Phase behaviour

We based our $n$-$p$-phase diagram (c.f. Fig. 1) on samples in the reference experiments. These samples crystallized in gas-tight bulk vials starting from a de-ionized and de-carbonized shear melt. Below the eutectic density, the phase diagram shows two broad ranges of substitutional alloys of bcc crystal structure to both sides of a small eutectic gap. This confirms previous observations on other eutectic mixtures [17, 21, 22, 59]. For all samples within the alloy regions, the characteristic microstructure is a polycrystalline solid of randomly oriented, intersection-facetted crystallites. Being completely solidified and being enclosed in gas-tight vials, these crystals are stable against microstructural transformations and melting for extended times. The same microstructure also appears during stage I in slit cells. Here, however, these early polycrystals dissolve and melt already during stage II and III. Also alloy crystals formed early in stage III dissolve again in favour of still later grown alloy crystals. Microstructural transformations ceased after some months. The final state was found to be a three-phase coexistence (c.f. Fig. 10a). In a homogeneous eutectic system this should only be possible exactly at the eutectic point. Here, however, we observe the phase behaviour in the presence of an external field. In fact, gravity establishes a region of very small $p$ at the very cell bottom, where β-phase is at local equilibrium with a coexisting fluid of $p < p_E$ [22]. Elsewhere the initial composition is more or less retained. There a fluid of $p > p_E$ should coexist in a local equilibrium with the α-phase. It is tempting to identify the refined late-stage alloy crystals grown on top of the β-phase (Fig. 8j) with this expected α-phase. The idea could be directly by an experimental determination of the composition of the alloy and the β-phase. Unfortunately, the present particles are too small for high resolution optical microscopy. However, future confocal experiments employing fluorescently labelled species could solve this issue [15].

Already now, however, we may draw another important conclusion. We recall, that during stage II and III the stage I alloy crystals and their sequels crystals melt in the absence of salt gradients. This also holds for the stage IV alloy crystals. Presumably, after five to seven months, local sedimentation equilibrium is attained and phase equilibrium established. Therefore, also the sequence during salt-induced melting reflects the relative thermodynamic stability of the different crystallite types (c.f. Fig. 11). From this, however, we may conclude, that the initially formed stage I alloys with $p = p_0$ (and most of their sequels grown early in stage III) are thermodynamically metastable with respect to the final small sphere enriched α-phase. Since bulk samples form stage I crystals only, this conjecture also applies to the there-observed polycrystalline solids. The crystalline regions in the phase diagram of Fig. 1b, therefore shows a region of mechanically stable but thermodynamically metastable alloy crystals. This possibility has been pointed out many years ago and in connection to polydisperse systems [23], but to the best of our knowledge has not been experimentally demonstrated for the case of mixtures. The true equilibrium tie-lines should thus boarder a much-extended fluid region. This suggestion could be supported e.g., by computer simulations using swap moves to overcome the slow transformation dynamics in solidified systems [27].

Stages of crystal formation

Our second point of discussion concerns the observed sequence of stages and its coupling to the different gradients. The four stages of crystal formation are schematically drawn in Fig. 11. Alloy formation in stage I proceeded without composition gradients, but in the presence of horizontally oriented, global gradients in electrolyte concentration and particle density. β-phase formation in stage II relied on a vertical gradient in $p$ induced by differential settling. The microstructural transformations of stage III were dominated by local, often lateral variations in $p$ at and around growing crystals. The final alloy formation stage IV was caused by large scale creaming of the smaller component and thus involved a global composition gradient.

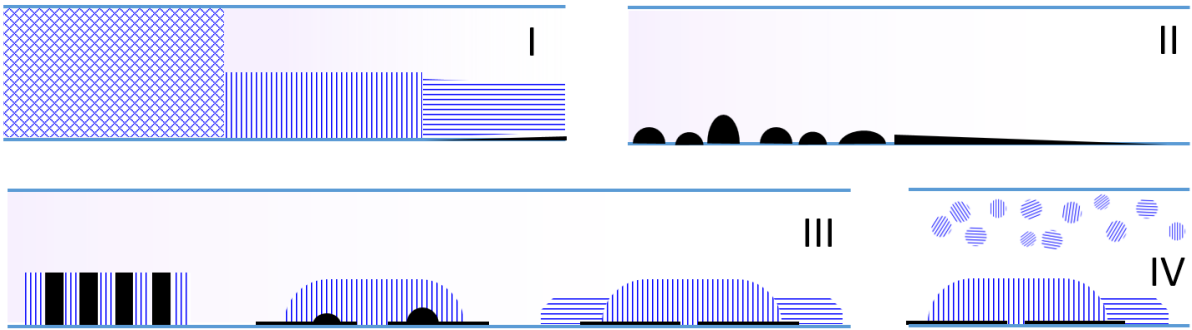

Figure 11: Schematic overview on the complex sequence of crystal growth. The mild purple background indicates the underlying density gradient. The initial salt-concentration gradient (not shown) vanishes already early in stage I. The four panels show the dominant crystal types of the four stages I – IV, respectively: Stage I is dominated by alloy crystal formation (hatched blue). This phase shows polycrystalline of randomly oriented crystallites (cross-hatched), vertical columns of crystals with preferred structural orientation (vertically hatched) and horizontal columns of random structural orientation (horizontally hatched). In stage II, sedimented large component particles form pebble-shaped and sheet-like β-phase crystals on the cell bottom (black). In stage III, a lager variety of crystallite morphologies emerges including (from left to right) sympathetic growth patterns of alloy- and β-phase as well as overgrowth of β-phase by alloy phase and secondary nucleated columnar alloy crystals. Earlier grown crystals dissolve in favour of later emerging ones. In stage IV, small particle enriched alloy crystals nucleate at random orientation in the upper cell part.

Both the sequence of stages and their duration were observed to be highly correlated to the characteristic time scales needed to develop the different gradients. Strength and overall lifetime of the stage I are set by the exchange rate and the cell geometry. The gradient advances quickly due to the high micro-ion diffusivity. The concentration profile evolves from initially step-like towards an eventually linear profile between the limiting concentrations set by exchange and leakage rate, respectively. For well-sealed cells, a nearly gradient-free, nearly fully de-carbonized state is reached within a week. Simultaneously, the gradient induced diffusiophoretic particle-transport establishes a particle-concentration gradient of opposite sign. I.e. low particle concentrations in the centre and enlarged particle concentrations in the reservoirs. Due to the much smaller diffusivity of the particles, this gradient is stable for extended times. The observed diffusiophoretic transport-velocities are on the order of cm/week. Unfortunately, the complex cell geometry, the presence of strong inter-particle repulsions, as well as the large uncertainties in the time-dependent relevant boundary conditions (exchange and leakage rates, local salt-concentration and salt-concentration gradient, $CO_2$-dependent $\zeta$-potential of the particles [51]), forbid any quantitative comparison to bulk theory [45, 60, 61] or full electro-hydrodynamic numerical modelling [62]. This interesting issue is therefore deferred to future investigations in simpler geometries and under well controlled local conditions.

In stage II the local vertical gradient in $p$ dominates. In the lowest part, $p$ soon falls below $p_E$ and with time it further decreases towards p = 0. This allows formation of $\zeta$-phase at the cell bottom. The formation time for this gradient is set by the ratio of sedimentation lengths and the observation region height [46]. It regularly occurs on the time scale of a few days. Incidentially, the ensuing β-phase formation also terminates further advance of the heterogeneous nucleation zone for stage I alloys. Formation of a global composition gradient takes much longer. In fact, the time scale for global creaming is set by the overall geometry of the cell. Creaming occurs within a few weeks. The corresponding increase in $p$ and/or $n$, , triggers stage IV alloy crystal formation (Fig. 4e and f; Fig. S6e in the SI).

In stage III local gradients in composition dominate the microstructural conversion processes. These were seen to result from expulsion of odd particles. The magnitude of local composition variation is coupled to the crystal growth velocities. Slow growth is more efficient for compositional refinement [63]. On the other side, fast growth may still yield a lot of expelled material. In our experiments we observe examples of both cases (c.f. Fig. 8c-f). The slow growing β-phase sheets and pebbles enrich small particles material on their upper surface and between them. Thus, the gradient in $p$ is locally disturbed. This is indirectly shown through the heterogeneous nucleation of stage III alloy crystals right next to growing pebbles (Fig. 5). More directly, the slow growing, late-stage central alloy crystals expel large particles very efficiently, increasing their purity and leading to a large stability against melting (Fig. 9 and 10a). Conversely, fast growing early-stage III columnar alloy crystals produce a quick pile-up of large particles triggering sympathetic growth of β-phase crystals (Fig. 4d), but are much less stable against melting. Like the stage I alloys, these structures vanish while further refined crystals still appear. In fact, several generations of refinement are needed to approach the final composition of stage III alloys. Local composition fluctuations thus hardly alter the general sequence but influence the duration of stage III. The presence of dominant gradients thus is seen to control the sequence and duration of the four observed crystallization stages.

Nucleation paths

A related third point of interest is the influence of gradients on the nucleation path taken within each stage. Here, it may be instructive to recall the dependence of nucleation paths on meta-stability known from previous investigations. There it was observed, that samples of decreasing metastability showed a transition from homogeneous nucleation to heterogeneous nucleation of wall-based columns to spherical caps and to thin crystalline monolayers [19, 53, 54, 64, 65]. The chosen nucleation scenario further depended on the specific shear history [50, 54, 55]. This in mind, we conclude, that the ordered array of stage I microstructures imaged in Fig. 5 must be due to an overall decrease in metastability. We rationalize this in a qualitative way as follows: At constant composition, metastability depends on both the salt and the particle concentration (c.f. Fig. 1b, 2d and $n$-$c$—phase diagrams in Fig. S1, of the SI). At the crystallization front, $c < c_F(n, p)$, $n > n_F(c, p)$, and $p \approx p_0$, where $p_0$ denotes the initial composition. Initially, the location of the front advances towards the cell centre, i.e. towards lower salt-concentrations. Due to the initial steepness of the gradient, the front is rather narrow. This results in an advancing burst of homogeneous nucleation and the formation of a fine grained Ia alloy microstructure characterized by straight crystallite intersections. As the formation of a density gradient sets in, the crystallization front slows and broadens. At the front location, the salt concentration further lowers with time, while now also the particle concentration decreases. In effect, these developments are counteracting regarding metastability. This, then leads to an extended, plateau-like region of moderate metastability. Within this region homogeneous nucleation and growth continue as a slow steady state process. In turn, this results in an extended region for stage Ib microstructure characterized by a comparably broad crystallite-size distribution and curved crystallite boundaries [58]. More towards the cell centre, the salt concentration has become constant while the particle density still decreases. A weak gradient in metastability results. Thus, we observe the expected transition to steady state heterogeneous nucleation of columnar crystals [53]. Recalling the literature, the next two stages in single component systems would actually be hemispherical caps and thin sheets [64, 65]. This, however is quenched in our two-component system by the competing formation of a thin layer of settled PnBAPS122. This layer impedes further heterogenous nucleation of alloys at the cell wall by lattice mismatch.

The expected transition is, however well visible for the β-phase in Fig. 6a. These β-phase pebbles and sheets form as the salt-concentration gradient has almost vanished, but the density gradient stays. The local metastability controls the degree of phase transformation. Larger metastability leads to pebbles, lower metastability to crystalline sheets. Due to ongoing differential settling, the sheet thickness increases slightly with time, but due to the gradient in $n$, it vanishes towards the cell centre. Interestingly, it is exactly there, where late-stage alloy crystals can nucleate again directly on the cell wall (c.f. Fig. 7, and 8b). Later on, thin sheets of PnBAPS122 may, however form in their vicinity from particles expelled during their crystal growth. This, presumably, creates a situation analogous to the transition from Ic to Id microstructures described above. It impedes further lateral growth of the alloy crystals and triggers secondary nucleation and outward growth of columnar crystals (Fig. 8d,f, and g).

Outlook

In the present study, we observed a wealth of microstructures. We found irregular, sheet-like and half-dome shaped crystals, crystals facetted by intersection, by growth kinetics and (presumably) also by surface tension effects. Single crystals as well as polycrystalline material was observed, and the applied shear history allowed further modifications. While giving a comprehensive overview, the present study had to remain qualitative. The next steps therefore should be experiments using differently dyed particles in order to quantify local concentrations and compositions, e.g. via fluorescence microscopy. Our study may also have contributed to a credible road map for aiming at specific microstructures. It is obvious, that the long-term goal remains a genuine microstructure-control for colloidal materials. The here observed relation of specific gradients and local metastability-conditions to characteristic microstructures may have aided that development.

On the other side, our study may have risen interest in a few novel phenomena. For instance, our observation of a gradual transition in the degree of faceting was correlated to the composition in the immediate vicinity. This poses the question of the existence and location of a roughening transition [66] for colloidal crystals. Further, sympathetic nucleation and growth were observable, but did not generally dominate as they would in metallic eutectics [1]. In fact, true eutectic lamellae remain to be demonstrated.

**Conclusions**

Our study demonstrated the metastability of polycrystalline alloys obtained from homogeneous shear-melts. This was found to underly a wealth of additional microstructures each forming at a specific combination of the key parameters $n$, $p$ and $c$. Throughout all slit cell experiments, the developing coupled gradients induced a spatial distribution of experimental conditions, to which the nucleation and growth processes reacted locally. The orthogonal orientation of the gradients with respect to each other allowed discriminating their respective influences. Moreover, we could follow their temporal evolution and compare the characteristic time scales of parameter variation and microstructural transformations. Thus, a reliable qualitative correlation of the local experimental conditions in $n$, $p$ and $c$ and the resulting microstructures became possible. Our approach clearly demonstrated the flexibility inherent in colloidal crystallization under internal gradients and opens novel paths towards specific microstructures not accessible in gradient-free solidification from homogeneous shear melts. We also anticipate, that our study may raise interest in the relation between colloidal crystallization and gradient induced (local and global) transport processes.

**Supporting material**

Further [supplementary material](#) on experimental details and additional data is available online.

**Acknowledgement**

We are pleased to thank Dieter Herlach, Patrick Wette and Enrique Vilanova Vidal for fruitful discussions. We thank Andreas Stipp, Max Hofmann and for providing images shot in Flow through cells. Financial Support of the DFG (Grant nos. Pa459/16 and Pa459/19-1), the Centre for INterdisciplinary Emerging MAterials, Mainz (CINEMA) and the JGU FoFö Stufe I (Project Metastable Solids) is gratefully acknowledged. I.G. held a DAAD Stipend within the IAESTE Program.

**Data Availability**

The raw data that support the findings of this study are available from the corresponding author upon reasonable request.

**Conflict of interest**

The authors have no conflicts to disclose

**Author Contribution**

Nina Lorenz devised and performed all measurements in slit cells and contributed a great part of measurements in the other cells. Ishtan Gupta contributed all measurements on the second mixture, shear modulus experiments and the SLS measurements. All authors contributed to data interpretation. Thomas Palberg wrote the manuscript with substantial contributions from Nina Lorenz.

**References**

1   H. Biloni, W. J. Boettinger, **Solification** in R. W. Cahn, P.Haasen (Ed.) *Physical Metallurgy 4th ed. Vol. I* North Holland, Amsterdam 1996

2   D. M. Herlach, *Euro. Phys. J. Special Topics* **223**, 591-608 (2014). **Colloids as model systems for metals and alloys: a case study of crystallization**

3   Y. Monovoukas, A. P. Gast, *J. Colloid Interface Sci.* **128**, 533-548 (1989). **The experimental phase diagramm of charged colloidal suspensions**

4   E. B. Sirota, H. D. Ou-Yang, S. K. Sinha, P. M. Chaikin, J. D. Axe, Y. Fujii, *Phys. Rev. Lett.* **62**, 1524-1527 (1989).**The complete phase diagram of a charged colloidal system: a synchrotron x-ray scattering study**

5   P. Wette, I. Klassen, D. Holland-Moritz, D. M. Herlach, H.J. Schöpe, N. Lorenz, H. Reiber, T. Palberg, S. V. Roth, *J. Chem. Phys.* **132**, 131102 (2010). **Complete description of re-entrant phase behaviour in a charge variable colloidal model system**

6   W. Luck, M. Klier, H. Wesslau, *Naturwissenschaften* **50**, 485-494 (1963). **Kristallisation übermolekularer Bausteine**

7   R. Niu, S. Heidt, R. Sreij, R. I. Dekker, M. Hofmann, T. Palberg *Sci. Rep.* **7**, 17044 (2017). **Formation of a transient amorphous solid formed from low density aqueous charged sphere suspensions**

8   B. J. Ackerson, N. A. Clark, *Phys. Rev. Lett.* **46**, 123-126 (1981). **Shear induced melting**

9   P. Tan, N. Xu, L. Xu, *Nature Phys.* **10**, 73–79 (2014). **Visualizing kinetic pathways of homogeneous nucleation in colloidal crystallization**

10  U. Gasser, *J. Phys.: Condens. Matter* **21**, 203101 (2009). **Crystallization in three- and two-dimensional colloidal suspensions**

11  M. Würth, J. Schwarz, F. Culis, P. Leiderer, T. Palberg, *Phys. Rev. E* **52**, 6415-6423 (1995). **Growth kinetics of body centred cubic colloidal crystals**

12  V. C. Martelozzo, A. B. Schofield, W. C. K. Poon, P. N. Pusey, *Phys. Rev. E* **66**, 021408 (2002). **Structural aging of crystals of hard-sphere colloid**

13  T. Palberg, *J. Phys.: Condens. Matter* **26**, 333101 (2014). **Crystallization kinetics of colloidal model suspensions: recent achievements and new perspectives**

14  E. V. Shevchenko, D. V. Talapin, N. A. Kotov, S. O'Brien, and Ch. B. Murray, *Nature* **439**. 55 - 59 (2006). **Structural diversity in binary nanoparticle superlattices**

15  E. C. M. Vermolen, A. Kuijk, L. C. Filion, M. Hermes, J. H. J. Thijssen, M. Dijkstra, and A. van Blaaderen, *Proc. Nat. Acad. Sci.* **106**, 16063-16067 (2009). **Fabrication of large binary colloidal crystals with a NaCl structure**

16  S. Hachisu, S. Yoshimura, Nature 283, 188 (1980). **Optical demonstration of superstructures in binary mixtures of latex globules**

17  T. Okubo, H. Fujita, *Colloid and Polymer Science*, **274**, 368 (1996). **Phase diagram of alloy crystal in the exhaustively deionized suspensions of binary mixtures of colloidal spheres**

18  P. Wette, H. J. Schöpe, T. Palberg, *Phys. Rev. E.* **80**, 021407 (2009). **Enhanced crystal stability in a binary mixture of charged colloidal spheres**

19  J. Liu, T. Shen, Z. H. Yang, S. Zhang, and G. Y. Sun, *J. Phys. Chem. Lett.* **8**, 4652-4658 (2017). **Multistep Heterogeneous Nucleation in Binary Mixtures of Charged colloidal Spheres**

20  A. Stipp, T. Palberg, *Phil. Mag. Lett.* **87**, 899-908 (2007). **Crystal Growth Kinetics in Binary Mixtures of Model Charged Sphere Colloids**

21  N. J. Lorenz, H. J. Schöpe, H. Reiber, T. Palberg, P. Wette, I. Klassen, D. Holland-Moritz, D. Herlach and T. Okubo, *J. Phys.: Condens. Matter* **21,** 464116 (2009). **Phase behaviour of deionized binary mixtures of charged colloidal spheres**

22  N. Lorenz, H. J. Schöpe, T. Palberg, *J. Chem. Phys.* **131**, 134501 (2009). **Phase behavior of a de-ionized binary charged sphere mixture in the presence of gravity** 23  J.L. Barrat et J.P. Hansen, *J. Phys. France* **47**, 1547-1553 (1986). **On the stability of polydisperse colloidal crystals**

24  P. B. Warren, *Phys. Chem. Chem. Phys.* **1**, 2197-2202 (1999). **Phase transition kinetics in polydisperse systems**

25  R. M. L. Evans, C. B. Holmes, *Phys. Rev. E* **64**, 011404 (2001). **Diffusive growth of polydisperse hard-sphere crystals**

26  T. Dasgupta, M. Dijkstra, *Soft Matter* **14**, 2465 - 2475 (2018). **Towards the colloidal Laves phase from binary hard-sphere mixtures via sedimentation**

27  P. K. Bommineni, N. R. Varela-Rosales, M. Klement, and M. Engel, *Phys. Rev. Lett.* **122** 128005 (2019). **Complex Crystals from Size-Disperse Spheres**

28  H. Löwen, A. van Blaaderen, J. K. G. Dhont, P. Nielaba, T. Palberg (Guest Ed.), *Euro. Phys. J. Special Topics* **222**, 2723-3087 (2013) **Colloidal dispersions in external fields**

29  C. P. Royall, E. M. Leunissen, A. van Blaaderen, *J. Phys.: Condens. Matter* **15**, 53381 (2003). **A new colloidal model system to study long-range interactions quantitatively in real space**

30  P. Dillmann, G. Maret, P. Keim, *Euro. Phys. J. Special Topics* **222**, 2941-2959 (2013). **Two-dimensional colloidal systems in time-dependent magnetic fields**

31  J. P. Hoogenboom, A. K. van langen-Suurling, J. Romijn, A. van Blaaderen, *Phys. Rev. E* **69**, 051602 (2004) **Epitaxial growth of a colloidal hard sphere hcp crystal and the effects of epitaxial (mis)match on the crystal structure**

32  B. J. Ackerson, S. E. Paulin, B. Johnson W. van Megen, S. Underwood, *Phys. Rev. E* **59**, 6903 (1999). **Crystallization by settling in suspensions of hard spheres**

33  M. E. Leunissen, M. T. Sullivan, P. M. Chaikin, and A. van Blaaderen, *J. Chem. Phys.* **128**, 164508 (2008). **Concentrating colloids with electric field gradients. I. Particle transport and growth mechanism of hard-sphere-like crystals in an electric bottle**

34  M. E. Leunissen and A. van Blaaderen, *J. Chem. Phys.* **128**, 164509 (2008). **Concentrating colloids with electric field gradients. II. Phase transitions and crystal buckling of long-ranged repulsive charged spheres in an electric bottle**

35  M. Trau, D. A. Saville, I. A. Aksay, *Langmuir* **13**, 6375-6381 (1997). **Assembly of colloidal crystals at electrode interfaces**

36  R. Niu, T. Palberg, *Soft Matter* **14** 3435-3442 (2018). **Seedless assembly of colloidal crystals by inverted micro-fluidic pumping.**

37  J. Palacci, B. Abécassis, C. Cottin-Bizonne, C. Ybert, and L. Bocquet, *Soft Matter* **8**, 980-994 (2010). **Osmotic traps for colloids and Macromolecules based on logarithmic sensing in salt taxis**

38  W. D. Dozier, P. M. Chaikin, *J. Phys. France* **43**, 843-851 (1982). **Periodic structures in colloidal crystals with oscillatory flow** 39  Y. L. Wu, D. Derks, A. van Blaaderen, and A. Imhof, *Proc. Nat. Acad. Sci.* **106** 10564-10569 (2009). **Melting and crystallization of colloidal hard-sphere suspensions under shear**

40  J. K. G. Dhont, M. P. Lettinga, Z. Dogic, T. A. J. Lenstra, H. Wang, S. Rathgeber, P. Carletto, L. Willner, H. Frielinghaus, P. Lindner, *Proc. Roy. Chem. Soc. Faraday Disc.* **123**, 157 – 172 (2003). **Shear banding and microstructure of colloids in shear flow**

41  Y. He, B. Olivier, B. J. Ackerson, *Langmuir* **13**, 1408 – 1412 (1997). **Morphology of crystals made of hard spheres.**

42  A. P. Gast, Y. Monovoukas, *Nature* **351**, 552-555 (1991). **A new growth instability in colloidal crystallization.**

43  L. Assoud, F. Ebert, P. Keim, R. Messina, G. Maret, and H. Löwen, *Phys.Rev. Lett.* **102**, 238301 (2009). **Ultrafast Quenching of Binary Colloidal Suspensions in an External Magnetic Field**

44  M. Leocmach, C. P. Royall, H. Tanaka, *Europhys. Lett.* **89**, 38006 (2010). **Novel zone formation due to interplay between sedimentation and phase ordering**

45  J. L. Anderson, *Ann. Rev. Fluid Mech.* **21**, 61-99 (1989). **Colloidal Transport by Interfacial Forces**

46  R. Piazza, *Rep. Prog. Phys.* **77** 056602 (2014). **Settled and unsettled issues in particle settling**

47  P. Wette, H.-J. Schöpe, R. Biehl, T. Palberg, *J. Chem. Phys.* **114**, 7556-7562 (2001). **Conductivity of deionised two-component colloidal suspensions**

48  T. Palberg, M. R. Maaroufi, A. Stipp and H. J. Schöpe, *J. Chem. Phys.* **137**, 094906 (2012). **Micro-structure evolution of wall based crystals after casting of model suspensions as obtained from Bragg microscopy**

49  Y. Monovoukas, G. G. Fuller, A. P. Gast, *J. Chem. Phys.* **93**, 8294 (1993). **Optical anisotropy in colloidal crystals**

50  M. Würth, J. Schwarz, F. Culis, P. Leiderer, T. Palberg, *Phys. Rev. E* **52**, 6415-6423 (1995). **Growth kinetics of body centred cubic colloidal crystals**

51  P. Vogel. N. Möller, M. N. Qaisrani, B. Pravash, S. Weber H.-J. Butt, B. Liebchen, M. Sulpizi, T. Palberg, *J. Am. Chem. Soc.* **144**, 21080-21087 (2022). **Charging of dielectric surfaces in contact with aqueous electrolyte – the influence of $CO_2$**

52  J. Yamanaka, N. Murai, Y. Iwayama, M. Yonese, K. Ito, T. Sawada, *J. Am. Chem. Soc.* **126** 7156-7157 (2004). **One-directional crystal growth in charged colloidal silica dispersions driven by the diffusion of base.**

53  J. Liu, H. J. Schöpe, T. Palberg, *J. Chem. Phys.* **116**, 5901-5907 (2002) and *ibid.* **123**, 169901 (2005) E. **Correlations between morphology, phase behaviour and pair interaction in soft sphere solid**s 54  M. R. Maaroufi, A. Stipp, T. Palberg, *Prog. Colloid Polym. Sci.* **110**, 83-88 (1998). **Growth and anisotropic ripening in twinned colloidal crystals**

55  A. Stipp, R. Biehl, Th. Preis, J. Liu, A. Barreira Fontecha, H. J. Schöpe, T. Palberg, *J. Phys.: Condens. Matter* **16** S3885-S3902 (2004). **Heterogeneous nucleation of colloidal melts under the influence of shearing fields**

56  H.-J. Schöpe, Th. Decker, T. Palberg, *J. Chem. Phys.* **109**, 10068-10074 (1998). **Response of the elastic properties of colloidal crystals to phase transitions and morphological changes**

57  Y. Monovoukas, A. P. Gast, *Langmuir* **7**, 460-468 (1991). **A Study of Colloidal Crystal Morphology and Orientation via Polarizing Microscopy**

58  W. A. Johnson, R. F. Mehl, *Trans. A. I. M. M. E.* **135**, 416 – 458 (1939). **Reaction kinetics in processes of nucleation and growth**

59  A. Meller, J. Stavans, *Phys. Rev. Lett.* **68**, 3646 - 3649 (1992). **Glass transition and phase diagramm of strongly interacting binary colloidal mixtures**

60  P. O. Staffeld, J. A. Quinn, *J. Colloid Interface Sci.* **130**, 69-87 (1989). **Diffusion induced banding of particles via diffusiophoresis: I. electrolytes**

61  P. O. Staffeld, J. A. Quinn, *J. Colloid Interface Sci.* **130**, 88-100 (1989). **Diffusion induced banding of particles via diffusiophoresis: II. non-electrolytes**

62  R. Niu, P. Kreissl, A. T. Brown, G. Rempfer, D. Botin, C. Holm, T. Palberg, J. de Graaf, *Soft Matter* **13**, 1505-1518 (2017). **Microfluidic Pumping by Micromolar Salt Concentrations.**

63  E. Tamborini, N. Ghofraniha, J. Oberdisse, L. Cipelletti, and L. Ramos, *Langmuir* **28**, 8562-8570 (2012). **Structure of Nanoparticles Embedded in Micellar Polycrystals**

64  P. Wette, A. Engelbrecht, R. Salh, I. Klassen, D. Menke, D. M. Herlach, S. V. Roth, and H. J. Schöpe, *J. Phys.: Condens. Matter*, **22**, 153101 (2009). **Competition between heterogeneous and homogeneous nucleation near a flat wall**

65  A. Engelbrecht, R. Meneses, H. J. Schöpe, *Soft Matter* **7**, 5685-5690 (2011). **Heterogeneous and homogeneous crystal nucleation in a colloidal model system of charged spheres at low metastabilities**

66  P. Nozieres, F. Pistolesi and S. Balibar, *Euro. Phys. J. B* **24**, 387 (2001). **Steps and facets at the surfaces of soft crystals**

Supplementary materials for

**Microstructural diversity, nucleation paths and phase behaviour in binary mixtures of charged colloidal spheres**

Nina Lorenz[1], Ishan Gupta[2], Thomas Palberg[1]

[1]Institute of Physics, Johannes Gutenberg University, 55122 Mainz, Germany

[2]Graz University of Technology, Institute of Applied Mechanics, Graz, Austria

Contents

A Materials

For the pure components, we determined the phase diagram measurements, following the observation procedures published initially by [67] and refined by Liu [68]. Samples were circuit conditioned (see below) and data taken in dependence on number density $n$ and concentration of NaCl, $c$. Starting from a thoroughly deionized sample in the crystalline state, we increased $c$ in steps and monitored the crystallization of the shear melt. Samples were observed by microscopy in a four-wall-polished flow through cell of rectangular cross-section. Crystallization end points were determined by observation at the narrow sides, microstructures were observed through the wide sides. Initially, only polycrystalline materials form. Closer to the phase boundary solidification proceeds via nucleation at the cell wall and subsequent growth. In the crystalline region of the phase diagram, wall crystals growing towards the centre meet, in the coexistence region, they don't. In the fluid region no crystals form. For PnBAPS70, we additionally performed a series of measurements under thoroughly deionized conditions, starting from the fluid state and increasing the number density.

Figure S1 a and b display the phase behaviour of PnBAPS70 and PnBAPS122, respectively. The inset in Fig. S1a gives a close-up of the lower end of the crystalline region. For clarity, we display the full data set only for the $n$-dependent measurement on PnBAPS70, with black triangles denoting crystalline samples, blue diamonds denoting samples in coexistence and red circles denoting fluid samples. Elsewhere, we only show the start and end of the determined coexistence region. The points given thus correspond to the freezing, respectively melting lines. The error bars denote the combined uncertainty in salt concentration and the location of melting and freezing. The error in number density as determined from the conductivity in the deionized state is about symbol size.

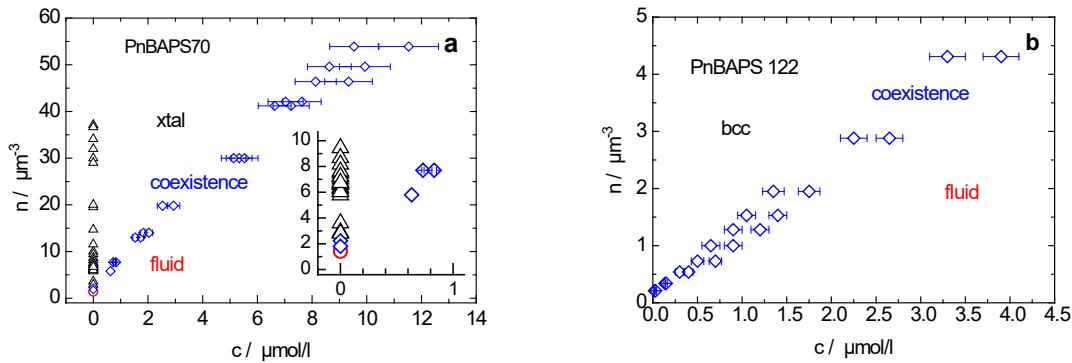

Fig. S1: Phase behaviour of the pure components in the $n$ / $c$ – plane: a) Phase diagram of circuit conditioned PnBAPS70 Symbols denote the phase state of the samples. Black triangles: crystalline; blue diamonds: upper and lower end of the coexistence region (melting and freezing); red circle: fluid. Inset: close-up of the low $n$, low $c$ region . At non-zero salt concentrations, we only show the coexistence region boundaries for clarity. b) Phase diagram of circuit conditioned PnBAPS122. Symbols as before.

As expected and in good agreement with previous work [69, 70, 71], we find a fluid phase at low number density and high salt concentration and a crystalline phase at large number density and low salt concentrations. The coexistence region is rather narrow and widens with increasing number density and salt concentration.

B Sample conditioning procedures

B1 Pre-Conditioning.

The original samples were shipped at a packing fraction of $\Phi \approx 0.2$. By dilution with distilled water, we prepared suspensions of approximately $\Phi = 0.1$, added mixed-bed ion exchange resin (IEX) (Amberlite UP 604, Rohm & Haas, France) and left them to stand under occasional stirring for some weeks. Then, the suspension was filtered using Millipore 0.5 µm filters to remove dust, ion-exchange debris and coagulate regularly occurring upon first contact of suspension with IEX. A second batch of carefully cleaned IEX filled into a dialysis bag was then added to retain low ionic strength in the stock suspensions now kept under Ar atmosphere in gas tight vials. All further preparation was done by either batch or circuit conditioning in three different cell types.

B2 Batch conditioning.

In batch conditioning, suitable amounts of the pre-conditioned pure species are filled under filtering into the sample cells containing some mixed bed IEX and diluted with milliQ water (Purelab Classic DI, ELGA, UK) to the desired concentration. Upon filling, samples get into contact with ambient air and take up $CO_2$, which dissociates and provides a finite background concentration of carbonic acid. The cells are then sealed with an air tight Teflon® septum screw cap, carefully avoiding air bubbles and then left to deionize.

B3 Circuit conditioning.

In circuit conditioning, suitable amounts of the pre-conditioned pure species are filled under filtering into a closed circuit system including the preparation units and the measuring cells [72, 73]. A schematic drawing is given in Fig. S4. The suspension is pumped by a peristaltic pump (P) through a closed and gas tight tubing system (black lines) connecting i) the ion exchange column (IEX); ii) a reservoir under inert gas atmosphere to add suspension, water or salt solutions (R); iii) a cell for conductivity measurements (C) and iv) one or more cells for the optical experiments (OC$_i$). Typically, one of these is a cylindrical cell (Quartz-cells of 10mm outer diameter, Hellma, Germany) for static light scattering, to determine the sample structure and/or number density. If desired, each actual measuring cells can be closed off from the circuit during measurements by additional electromagnetic valves (not shown). Optionally, also an online filter can be integrated past the IEX column to remove IEX debris and occasional shear coagulate (mainly doublets [74]).

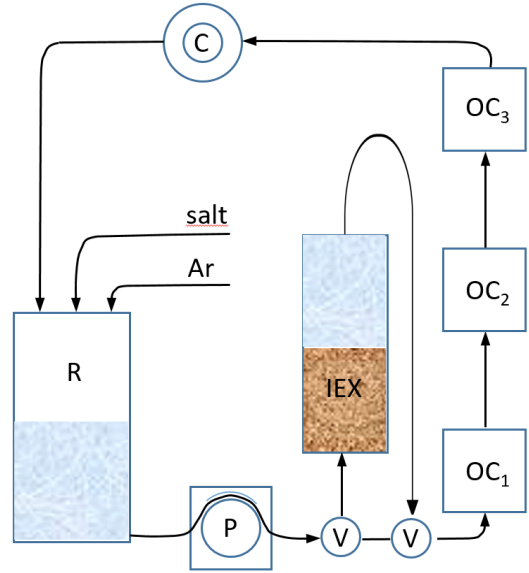

Figure S2: Conditioning circuit. Letters denote: R: reservoir; P: peristaltic pump; V: three way valves. IEX: mixed bed ion exchange column; OC$_1$, OC$_2$, OC$_3$, … optical flow through cells; C: conductometric cell; Additional electromagnetic valves can be introduced before and after measuring cells. All components are connected by gas-tight Teflon® tubings, with flow direction indicated as arrows.

The main advantages of circuit conditioning are the well-controlled experimental boundary conditions and the option to apply directed shear in the observation cell prior to crystallization. Care is taken to assure a circuit free of larger $CO_2$-leaks. All components are connected by Teflon® tubing, except for the short part at the pump, which is made from silicone rubber. Connection is made by gas tight tube fittings and three way valves (V), either custom made or commercial (Bohländer, Germany). Conductivity is measured at a frequency of $\omega = 400$ Hz [electrodes LTA01 and LR325/01 with bridge LF538 or electrode LR325/001 with bridge LF340, WTW, Germany]. To check for reproducibility we compared conductivity values at different frequencies $\omega < 1$kHz but found no dependence on $\omega$. Thus $\omega$ is low enough to measure the DC-limit of conductivity but large enough to inhibit significant electrophoretic particle transport. Care was further taken to control the suspension temperature within 1K. In general the reproducibility of conductivity measurements in suspensions was found to be better than 2%. The conductivity of deionized samples is proportional to the number density, $n$, and the effective conductivity charge, $Z_\sigma$, corresponding to the effective number of freely moving counter-ions [75, 76].

Typically, the suspension is cycled until a stable minimum conductivity is reached, defining the thoroughly deionized state. This takes about one hour. The minimum residual ion concentration is then given from the ion product of water: $[H^+][OH^-] = 10^{-14}$ mol$^2$L$^{-2}$. The proton concentration, $[H^+] = n\, Z_\sigma\, /\, 1000\, N_A$ is set by the amount of counter-ions released by the particles and amounts to a few $\mu$molL$^{-1}$. $N_A$ is Avogadro's number. After thorough deionization, the ion exchange chamber is bypassed. Circuit preparation provides homogeneous, thoroughly deionized suspensions. Deionized conditions are stable for a few hours, before small leaks (e.g. mainly through the silicone part of the tubing) lead to a noticeable rise in ion concentration. For measurements at elevated salt concentration, the suspensions are first deionized. Then NaCl (Titriplex, Merck) is added to the reservoir and the system homogenized again under bypassing the IEX column. To infer the resulting salt concentration from the conductivity reading we use the known $Z_\sigma$ in Hessingers model of independent ion migration [Hessinger]

Circuit conditioning was employed for PnBAPS112-PS392 phase diagram measurements in flow through cells of rectangular cross section of 1mm × 10mm and 2mm × 10mm, respectively [Rank Bros. Bottisham, UK or Lightpath Optical, UK]. It also was occasionally employed in slit cells, integrated via suitable gas tight fittings.

B4 Cylindrical vials.

Cylindrical 2ml glass vials (Supelco, Bellefonte, PA, USA) are suitable only for batch conditioning. After filling and sealing, they are placed cap down on a shelf. Under occasional gentle shaking, they were exhaustively deionized over several weeks. The electrolyte concentration cannot be directly measured in batch conditioning. Rather, the typical time to reach exhaustive deionization was determined indirectly with samples at large $n$. There, a polycrystalline microstructure formed reproducibly after shaking and re-crystallization. Shaking transforms existing crystals into a homogenized shear melt. The homogeneous nucleation rate increases roughly exponentially with increasing metastability. Larger nucleation rates yield smaller crystals. The constancy of crystallite radii in polycrystalline samples indicated thorough deionization. For samples at larger $n$, it took about one month to reach the thoroughly deionized state. If desired, the crystal structure, lattice constant and number density are then determined from static light scattering [Schöpe multipurpose].

B5 Slit cells.

Slit cells (Micro Life®, Hecht, Berlin) are well suited for slow batch deionization. Figure S3a shows an empty Micro Life® cell in upright position. These feature two cylindrical reservoirs of volume V ≈ 1.5ml connected by a thin parallel plate observation slit. The slit spans 47mm in x-direction (between reservoir rims), has a width of $y \approx 7.5$mm and a height of $z = 500$μm. About 0.75ml of IEX is filled in each reservoir. The observation chamber is IEX-free. After filling in the suspension, the cell openings are sealed by screw caps with Teflon® septum and left for deionization, typically in cap-down position. Alternatively, they can be transiently integrated into circuit conditioning by custom made screw connections. After conditioning, they are disconnected and the connections replaced again by the screw caps. While the Teflon® septum caps are gas-tight, the glue used in cell fabrication is not. Therefore, we additionally sealed the cell at all its seams. In Figure S3, the applied epoxy glue is seen as orange material. This drastically lowered the leakage rate. Only in rare cases, larger leaks are left. They can be easily identified by strictly local melting.

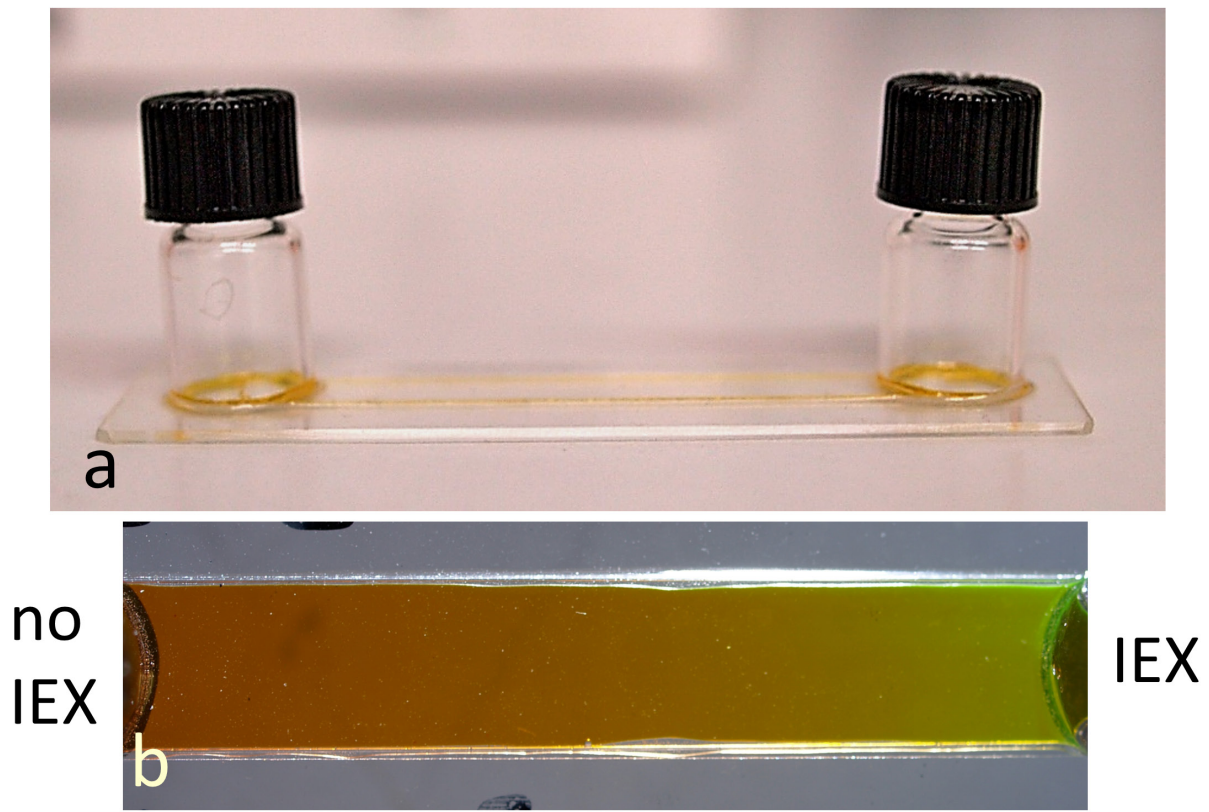

Figure S3. a) Slit cell in upright position before filling. Note the epoxy resin sealed seams. b) Gradient in salt concentration evolving in a slit cell with IEX placed only at one side. Image taken about 1h after filling. The gradient is visualized using 0.05molL$^{-1}$ aqueous Uranine solution which changes colour in dependence on concentration in a calibrated way. In the green region the concentration has fallen to c ≈ 0.01molL$^{-1}$

The local salt concentration cannot be determined directly. However, we demonstrate the evolution of a salt concentration gradient in Fig. S3b. Here we placed IEX in only one of the reservoirs and filled the cell with 0.05molL$^{-1}$ of Uranine solution [Merck, Germany]. Uranine solutions change their colour in dependence on concentration. This was calibrated spectroscopically (not shown). Starting from an initially homogeneous salt concentration, a pronounced gradient evolves as seen by the lateral colour change.

With sufficient sealing, a constant very low electrolyte concentration in slit cells is reached under batch conditioning in about one month. In principle, deionized conditions should be maintained over very long times. However, even the epoxy glue is not perfectly gas tight. The electrolyte concentration, is kept very low by continuing ion exchange. The IEX becomes exhausted after 7-11 months a year (depending on its amount compared to leak rate), and the salt concentration slowly rises again. The increase is sufficiently slow to allow for diffusive distribution of carbonic acid throughout the cell and an efficient levelling of gradients. For our melting experiments, we assume the electrolyte concentration to be practically homogeneous.

C Additional microstructures

C1 Microstructures observed in batch conditioned cylindrical vials:

Examples of fully polycrystalline microstructures were shown in Figure 1a of the main text. At large $n$, these are observable shortly after last shaking. At low n, reaching the final microstructure after last shaking may take much longer. In Figure S4a-e, we show samples at different deionization stages and different times after last shaking. In a), alloy crystallites nucleated and grew throughout, then settled under gravity. In b), PnBAPS70 enriched crystals nucleated throughout, but then creamed towards the top forming a so-called "head crystals" [77]. In c), we observe crystal formation by heterogeneous nucleation and growth on the IEX and at the cell walls. Columnar microstructures prevail. Figure S4d and S4e, we show examples with mixed microstructures, where the determination of a fully crystalline state is difficult. In particular, the sample in Fig. S4e appears to be filled only with a loose piling of homogeneously nucleated crystals.

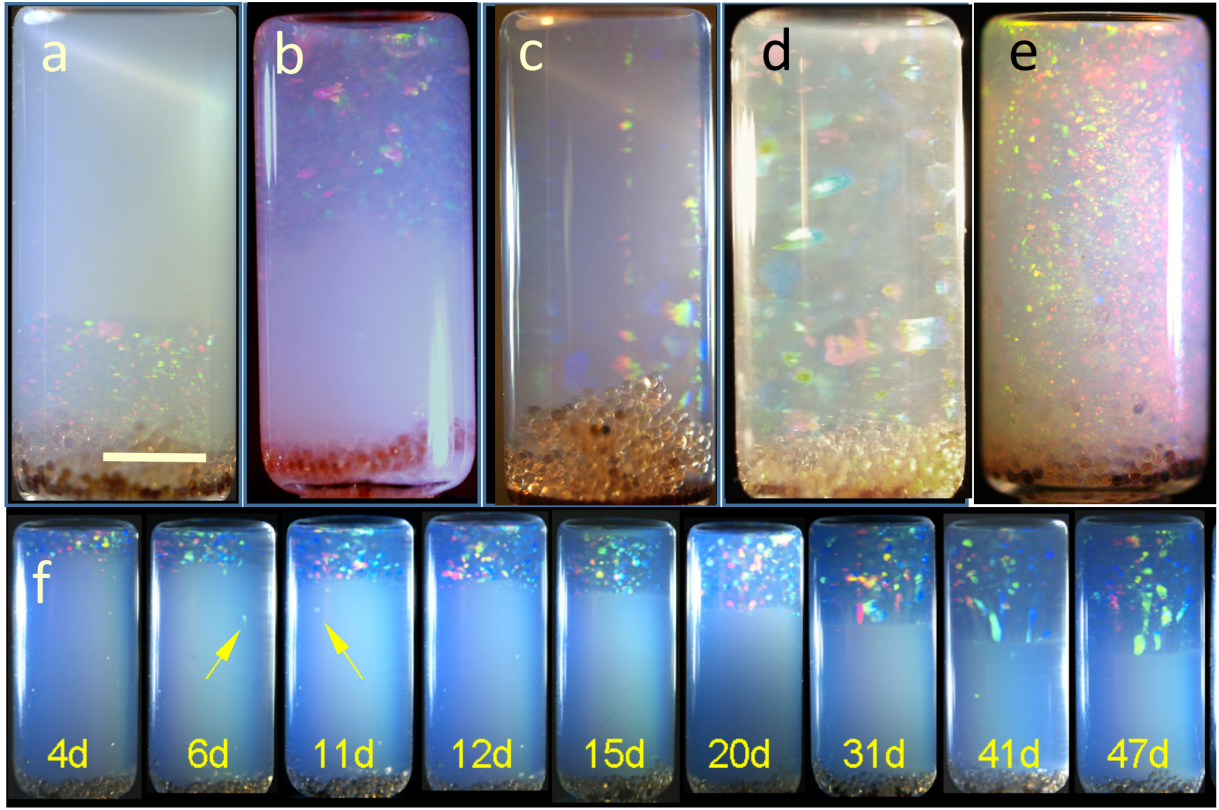

Figure S4: Example photographic images of samples obtained by batch deionization in bulk vials. The slightly differing overall colour originates from different illumination conditions. Note the differences in observable microstructures. Scale bar 500μm. a) Sample at $p = 0.9$ and $n = 14.2$μm$^{-3}$ last shaken after one week of deionization and imaged 1d later. b) Sample at $p = 0.96$ and $n = 18$μm$^{-3}$ last shaken after two weeks of deionization and imaged two weeks later. c) Sample at $p = 0.972$ and $n = 6.8$μm$^{-3}$ last shaken after five weeks of deionization and imaged three weeks later. d) Sample at $p = 0.94$ and n = $10.2$μm$^{-3}$ last shaken after five weeks of deionization and imaged 2d later. e) Sample at $p = 0.877$ and n = $32$μm$^{-3}$ last shaken after three weeks of deionization and imaged 1d later. f) Sample at $p = 0.90$ at n = $12$μm$^{-3}$ last shaken after five weeks of deionization and imaged at different additional times thereafter, as indicated.

The growth of "head crystals" became well reproducible after thorough deionization. These nucleate very slowly in the top half of the vials. Reaching a final stable microstructure may take up several weeks. In Figure S2f we show a thoroughly deionized sample at $p = 0.90$ at n = $12$μm$^{-3}$ imaged at different times after last shaking. Different to the slit cell stage IV crystals, the vial grown crystals formed a compact polycrystalline solid by further creaming of individual crystals (arrows). After about three weeks the solidification path changed and the sample showed a downward growth of columnar crystals commencing at the lower surface of "head crystals". Static light scattering yielded $p = 0.93 \pm 0.05$ and $n = 12.5$μm$^{-3}$ for the creamed polycrystals and $p = 0.79 \pm 0.05$ and n = $9.8$μm$^{-3}$ for the fluid region.

C2 Microstructures in circuit conditioned slit cells

In circuit conditioned slit cells, nearly all samples, except those very close to or in the coexistence of the bulk phase diagram, crystallized completely. At low $n$, we obtained polycrystalline samples. At intermediate $n$ we sometimes observed elongated twin structures. At large $n$ we obtained a characteristic fuzzy microstructure of near uniform colour, indicating a preferred crystallite orientation, presumably related to the high shear rates present during circulation and very fast crystallization. Two examples are shown in Figure S5 a and b.

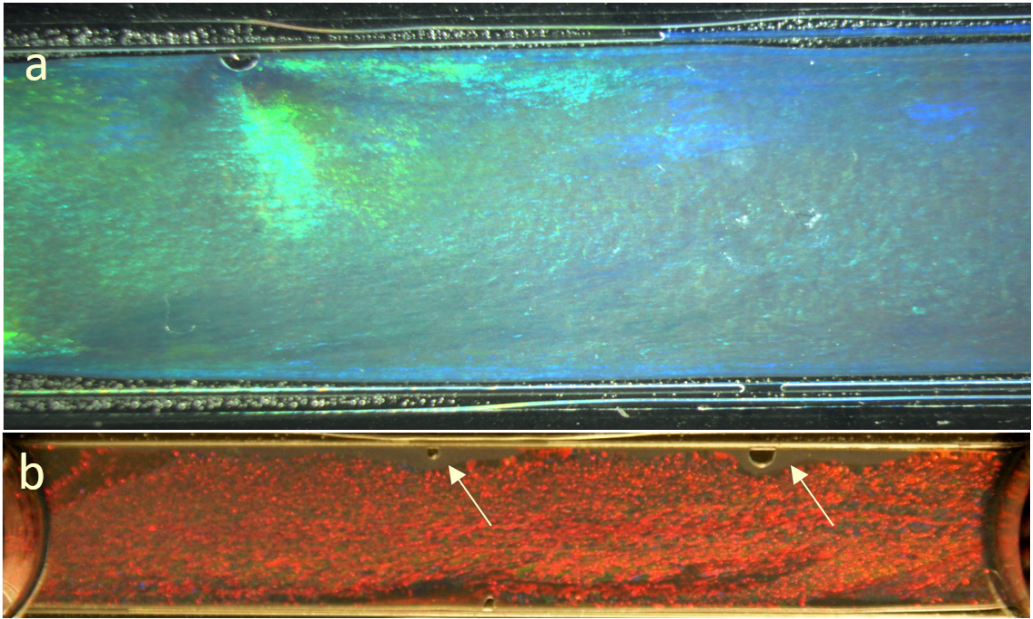

Figure S5: Microstructures observed after circuit conditioning in slit cells. a) Bragg image of a sample at $p= 0.96$ and $n = 32 \mu m^{-3}$. h) Bragg image of a sample at $n = 100 \mu m^{-3}$ taken 2d after circuit conditioning. Arrows mark regions molten due to $CO_2$-leakage.

These microstructure are remarkably stable against any evolution. Neither any significant coarsening nor formation of any of the later microstructural types discussed in the main text is seen. Possibly, evolution is hindered by the vertical confinement. Melting starts at the cell seams. Early local melting occurred at larger leaks (arrows in Fig. S5b).

## C3 Microstructures in batch conditioned slit-cells

Fig. S6 presents large scale images of two batch.conditioned slit cell samples.

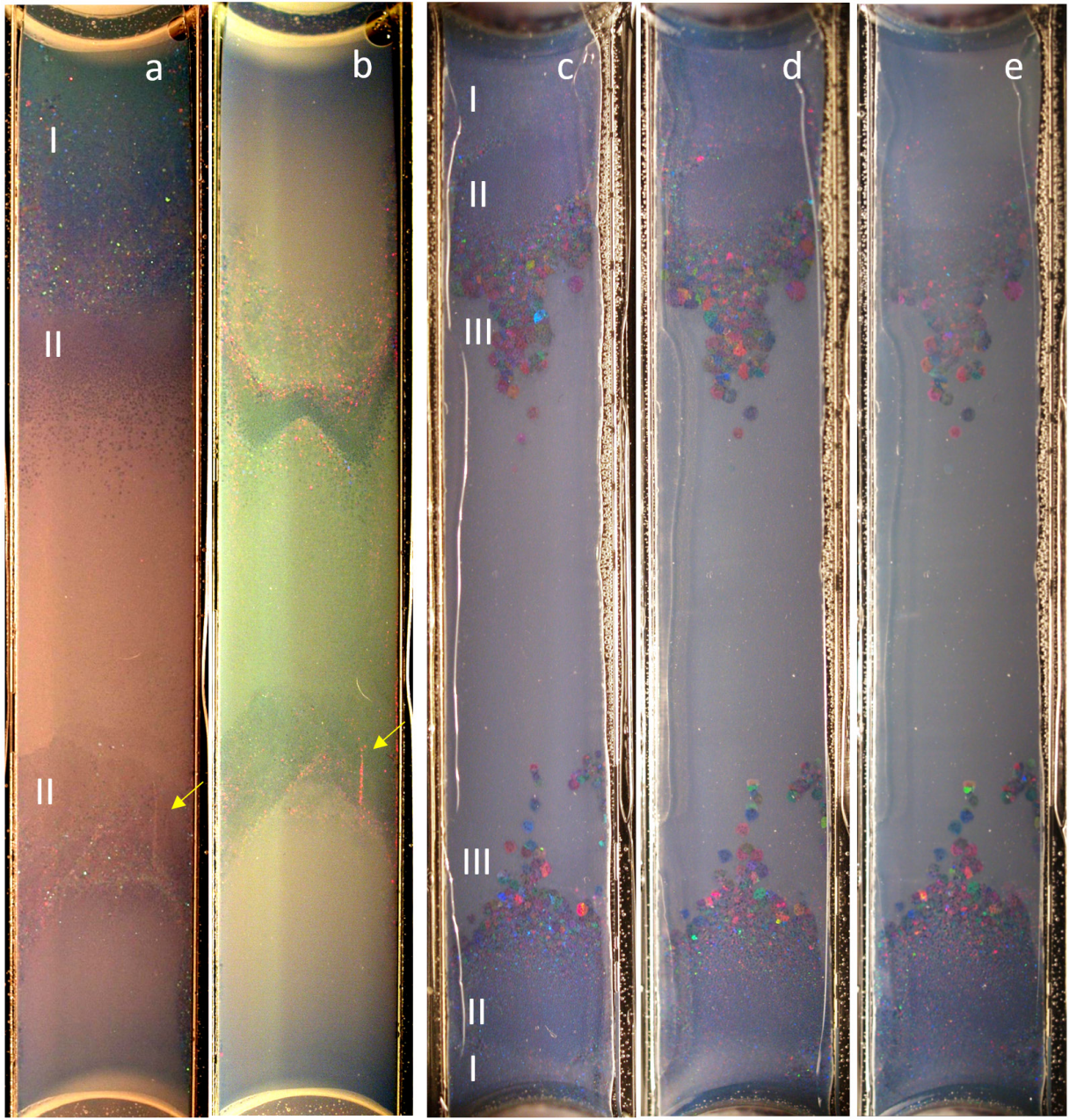

Fig. S6: Images of PnBAPS70/PnBAPS122 in slit cells. Image height 51.8mm, image width 7.8mm. Regions of crystals formed during different phases are marked by roman numerals. a) $p = 0.22$, $n = 5.6\mu m^{-3}$, $t = 6d$. Initial diffusiophoretic gradient with lower particle concentration in the centre. The arrow marks a specific local feature. b) The same sample as in a) at $t = 35d$. Note the shifted feature and the inversed gradient due to creaming of PnBAPS70. c) to d) Images taken on a sample at $p = 0.90$ and $n = 47\mu m^{-3}$ in an optimally sealed cell in cap-down orientation. Images taken at $t = 8d$, $t = 17d$ and $t = 28d$ after filling, respectively. Note the homogeneous colouring of the fluid and the stationarity of the crystals.

From filling, all sample contain $CO_2$, which dissociates to form carbonic acid. At the IEX, $CO_3H^-$ is exchanged for $OH^-$. This establishes a salt concentration gradient. Throughout the bulk fluid, a diffusiophoretic drift of negative particles results, along the $CO_3H^-$ gradient, towards the IEX [78, 79]. This may result in particle concentration gradients, which may become rather pronounced at lower number densities. Figure S5a and b show a time series of a sample at $n = 12.8 \mu m^{-3}$ and $p = 0.5$.

The sample was Bragg imaged under oblique white light illumination with the angle chosen suitably to capture the maximum of the static structure factor of the fluid. At fixed illumination angle, the ordered fluid regions display a homogeneous colour depending on particle concentration. Red indicates a lower particle concentration than green. Embedded crystals show vivid Bragg reflections, if oriented favourably. Bottom based β-phase is not oriented favourably and appears as black spots or films, respectively. In Figure S6a, six days of diffusiophoretic transport in the salt gradient have diluted the central region of the cell, resulting in a reddish diffraction colour. However, after about three weeks the drift direction reversed. In the central region, the systems became even more concentrated than before (Figure S5b). This effect is due to creaming of PnBAPS70. Creaming is more pronounced in low-$p$ samples. Here, the melt-$p$ presumably stays below $p_E$, such that no stage IV crystals can nucleate homogeneously. The process thus can continue much longer and the density increase further.

Creaming is much less pronounced at large $p$ and may even be missing. An example is shown in Fig. SS6c to S6e. Creaming did not occur, even though stage I alloys formed only out to the rims of the observation region, and the central fluid region stayed very large. The image series further shows one of the rare cases of an optimally sealed cell. In this high-density sample we observe neither creaming induced nor diffusiophoretically induced density gradients. No drift is observable for the β-phase region and the facetted-alloy crystals.

Due to the sketched transport processes, samples often show pronouncedly curved growth fronts and boundaries between different microstructures. Fig. S7 gives an example. As in the main text, the different microstructures are labelled by lower case letters. To the left (a), we see fine grained polycrystalline material formed after an initial nucleation burst. More to the right (b), we find larger grains with curved edges formed during subsequent steady state nucleation. The rightward edge is populated by columnar crystals heterogeneously nucleated at the rightmost type b crystals and growing sideways (c). The intermediate crystal type (c; heterogeneously nucleated at the cell bottom) has not developed in this case.

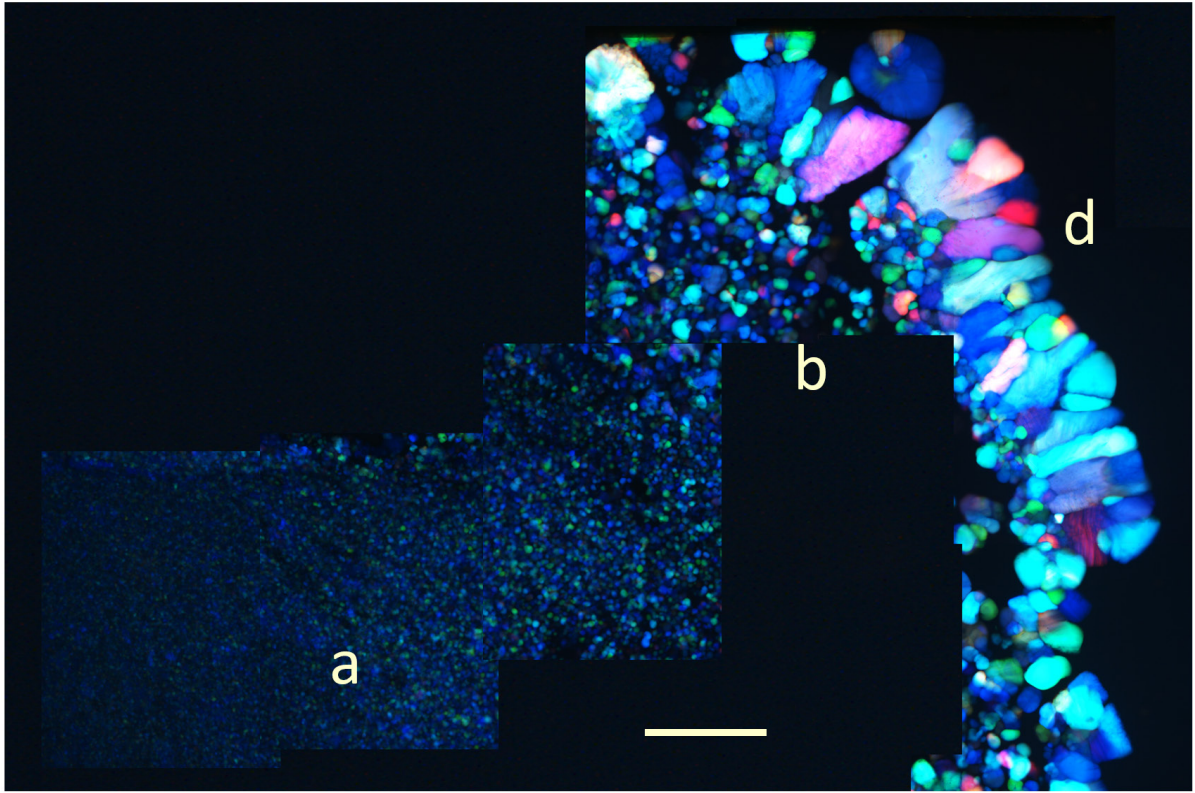

Fig. S7: Collage of PM micrographs showing part of a curved stage I alloy crystal growth front. Scale bar 500µm. Image taken at $t = 6$d on a sample with $p = 0.94$ and n = 32µm$^{-3}$. Lower case letters denote different microstructures of these stage I alloy crystals as defined in the main text.

As stage III progresses, more and more alloy crystals form in the central region on top and aside the β-phase sheets or (in the most central regions) even on the bare substrate. Fig. S8 shows an enlarged high resolution PM collage of the sample shown also in Fig. 6 of the main text. Note the pronounced internal texturing of the crystals.

Interestingly, systems at the same composition of $p = 0.94$ but prepared at larger density of $n = 20$ µm$^{-3}$, show an overall similar behaviour in stage II and III but quite some difference in detail. This is shown in Fig. S9.

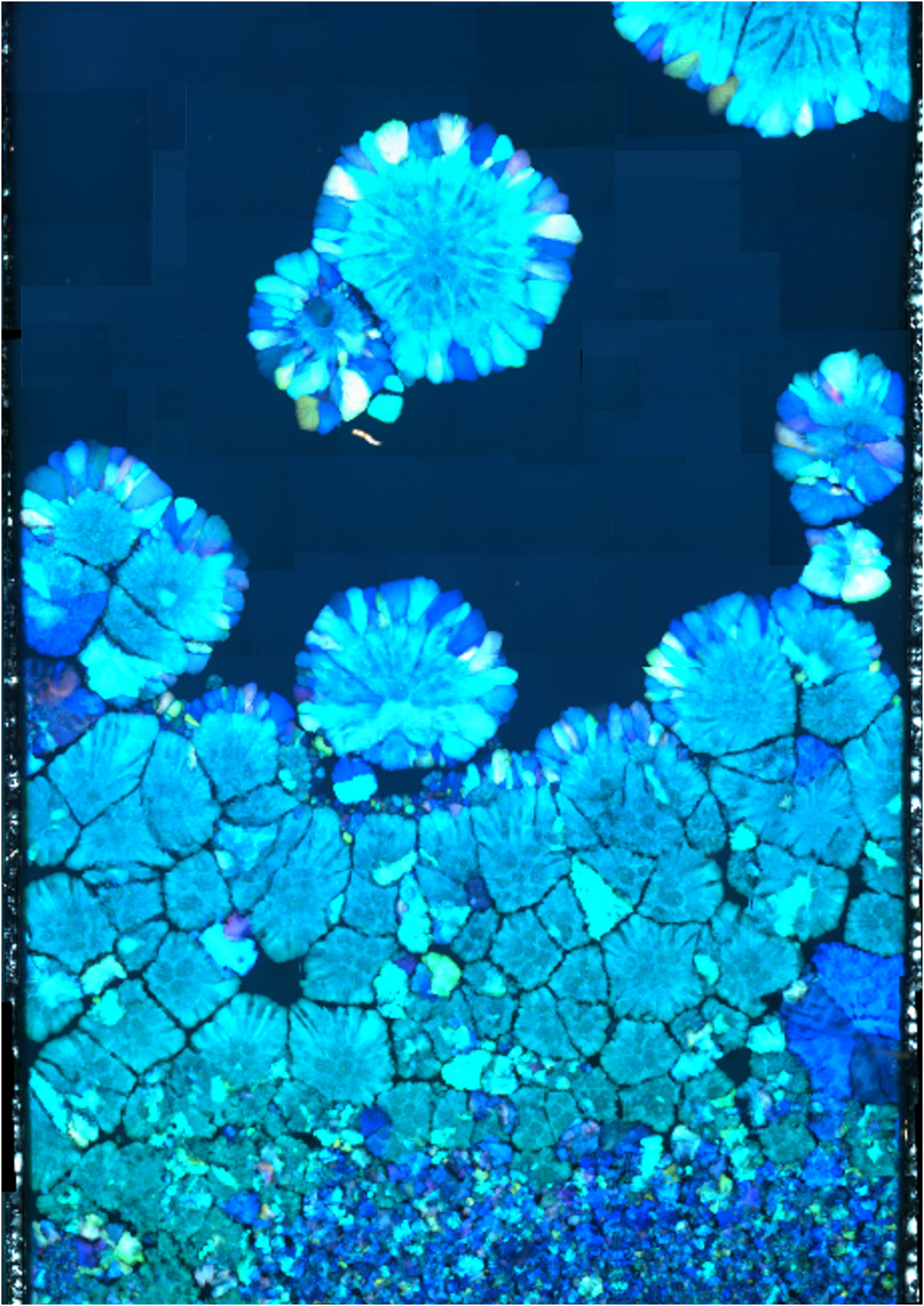

Figure S8: high resolution PM collage of a sample at $p = 0.94$, $n = 15\mu m^{-3}$ and $t = 50d$.

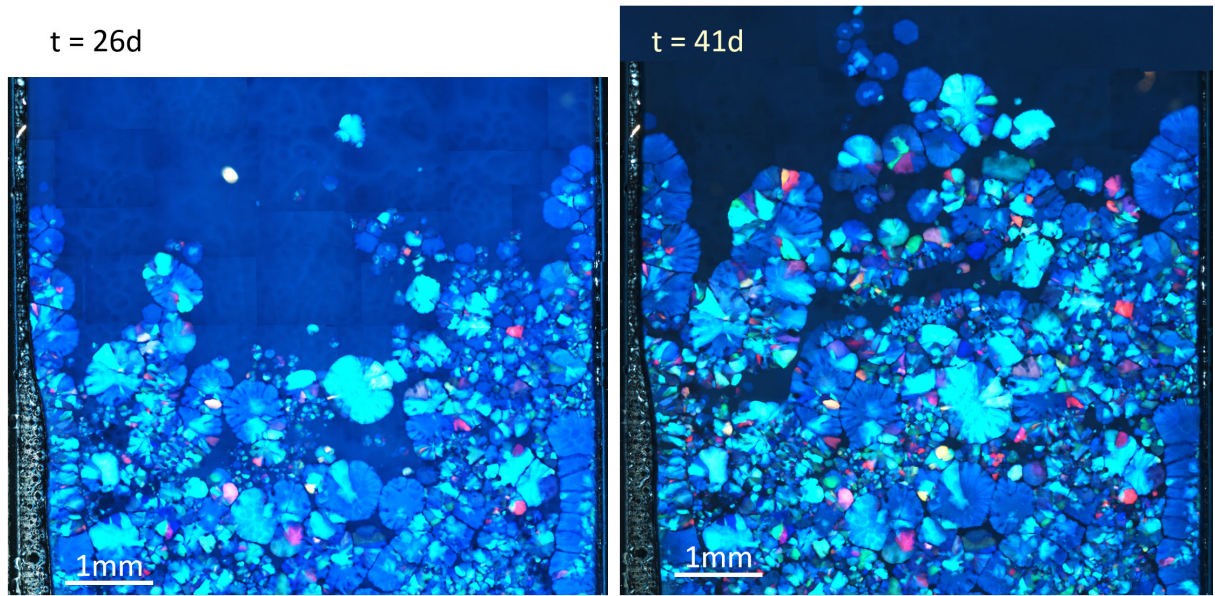

Figure S9: Collages of PM micrographs taken on a sample at at $p = 0.94$, $n = 20\mu m^{-3}$ and times as indicated. Stage III alloy crystals form in the central region on top and aside more highly grown β-phase crystals. The left panel is displayed with enhanced brightness to highlight the initially underlying β-phase microstructure.

As compared to Fig. S8, the central β-phase has developed more and higher patches but hardly sheets. The stage III alloy crystals again grow on top and/or overgrow the latter. We now observe a broad dispersion of crystal sizes with many more small and randomly oriented crystals and much less larger crystals with pronounced growth faceting. In addition, we occasionally observe randomly oriented alloy crystals on top of uniformly oriented larger alloy crystals. At still larger $n$, this crystallization type becomes even more pronounced frequent (see Fig. 7h of the main text).

Also the character of β-phase crystals differs markedly from that seen at lower $n$. We now observe a large number of enclosed near circular pebbles (see also Fig. 7c of the main text), longish enclosures emanating during outward growth of the larger alloy crystals (see also Fig. 7e and f of the main text) and irregular polygons embedded at the intersection of the smaller ones (as in Fig. 5d of the main text). Where it is still flat, the β-phase patches are easily overgrown, as highlighted in the PM micrograph displayed in Fig. S10.

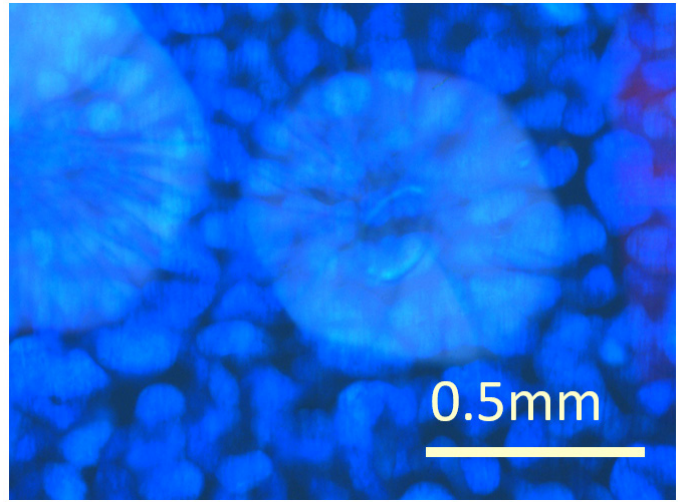

Figure S10: PM Micrograph of a central alloy crystal, overgrowing a bottom based, dense assembly of very thin β-phase patches.

Some of the late stage crystals developed a pronounced faceting. We analysed it by image processing. Fig. S10 gives a sketch of the measurements performed. For the majority of cases, the orientation and dimension of facets was compatible with either the symmetry of the (110) plane of bcc crystals or of the (111) plane of fcc crystals. From studies on ca. 80 crystals investigated over time, we observed a trend for the fraction of rectangular crystals. Their relative fraction increased with their nucleation date. Less hexagonal crystals formed at later stages.

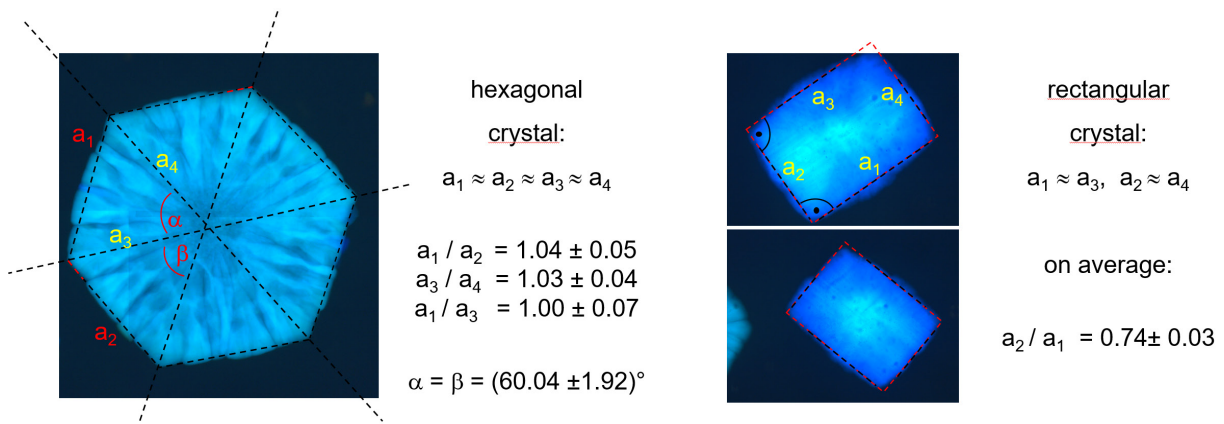

hexagonal crystal:

$a_1 \approx a_2 \approx a_3 \approx a_4$

$a_1 / a_2 = 1.04 \pm 0.05$
$a_3 / a_4 = 1.03 \pm 0.04$
$a_1 / a_3 = 1.00 \pm 0.07$

$\alpha = \beta = (60.04 \pm 1.92)°$

rectangular crystal:

$a_1 \approx a_3, \ a_2 \approx a_4$

on average:

$a_2 / a_1 = 0.74 \pm 0.03$

Fig. S10 sketch of the image analysis performed on growth facetted crystals. The hexagonal crystal had grown for 18d, the rectangular one for 24d.

References

67	M. Würth, J. Schwarz, F. Culis, P. Leiderer, T. Palberg, *Phys. Rev. E* **52**, 6415-6423 (1995). **Growth kinetics of body centred cubic colloidal crystals**

68	J. Liu, T. Palberg, *Prog. Colloid Polym. Sci.* **123**, 222-226 (2004). **Crystal growth and crystal morphology of charged colloidal binary mixtures**

69	E. B. Sirota, H. D. Ou-Yang, S. K. Sinha, P. M. Chaikin, J. D. Axe, Y. Fujii, *Phys. Rev. Lett.* **62**, 1524-1527 (1989). **The complete phase diagram of a charged colloidal system: a synchrotron x-ray scattering study**

70	S. Alexander, P. M. Chaikin, P. Grant, G. J. Morales, P. Pincus, D. Hone, *J. Chem. Phys.* **80**, 5776-5781 (1984). **Charge renormalization, osmotic pressure, and bulk modulus of colloidal crystals: Theory**

71	P. Wette, I. Klassen, D. Holland-Moritz, D. M. Herlach, H.J. Schöpe, N. Lorenz, H. Reiber, T. Palberg, S. V. Roth, *J. Chem. Phys.* **132**, 131102 (2010). **Complete description of re-entrant phase behaviour in a charge variable colloidal model system**

72	T. Palberg, W. Härtl, U. Wittig, H. Versmold, M. Würth, E. Simnacher, *J. Phys. Chem.* **96**, 8180-8183 (1992). **Continuous deionization of latex suspensions**

73	P. Wette, H.-J. Schöpe, R. Biehl, T. Palberg, *J. Chem. Phys.* **114**, 7556-7562 (2001). **Conductivity of deionised two-component colloidal suspensions**

74	J. Schwarz, P. Leiderer, T. Palberg, *Phys. Rev. E.* **104**, 064607 (2021). **Peaked bulk crystal nucleation in charged sphere melts from salt concentration dependent crystallization experiments at very low metastability**

75	L. Shapran, H. J. Schöpe, T. Palberg, *J. Chem. Phys.* **125**, 194714 (2006). **Effective charges along the melting line of colloidal crystals**

76	L. Shapran, M. Medebach, P. Wette, H. J. Schöpe, T. Palberg, J. Horbach, T. Kreer, A. Chaterji, *Colloid. Surf. A* **270**, 220-225 (2005). **Qualitative characterisation of effective interactions of charged spheres on different levels of organisation using Alexander´s renormalised charge as reference**

77	N. Lorenz, H. J. Schöpe, T. Palberg, *J. Chem. Phys.* **131**, 134501 (2009). **Phase behavior of a de-ionized binary charged sphere mixture in the presence of gravity**

78	J. L. Anderson, *Ann. Rev. Fluid Mech.* **21**, 61-99 (1989). **Colloidal Transport by Interfacial Forces**

79     T. Palberg, M. Würth, *Phys. Rev. Lett.* **72**, 786 (1994). **Comment on Tata et al.:„Vapour-liquid condensation in charged colloidal suspensions**